\documentclass[a4paper,11pt]{article}
\usepackage{jheppub} 
\usepackage[utf8]{inputenc}
\usepackage[T1]{fontenc} 
\usepackage{amsmath,amsthm,amssymb}
\usepackage{graphicx}
\usepackage{subfigure}
\usepackage{xcolor}
\usepackage{float}
\usepackage{array}
\usepackage{arydshln}
\graphicspath{{fig/}{image/}}
\usepackage{hyperref}
\hypersetup{colorlinks=true,linkcolor=blue,citecolor=magenta}

\newcommand{\Chat}{\widehat{\mathcal{C}}}

\newcommand{\dd}{\mathrm{d}}

\newtheorem*{thm*}{Theorem}

\title{\boldmath Linear stability matrix and Lyapunov  exponents for stationary spacetime}

\author[a,b]{Xu Ye}
\author[1,a,b]{and Shao-Wen Wei,\note{Corresponding author.}}

\affiliation[a]{Lanzhou Center for Theoretical Physics, Key Laboratory of Theoretical Physics of Gansu Province, Key Laboratory of Quantum Theory and Applications of MoE, Lanzhou University, Lanzhou 730000, China}
\affiliation[b]{Institute of Theoretical Physics \& Research Center of Gravitation, School of Physical Science and
	Technology, Lanzhou University, Lanzhou 730000, China}

\emailAdd{yex2026@lzu.edu.cn}
\emailAdd{weishw@lzu.edu.cn}

\abstract{
	Lyapunov exponents characterize the average rate at which nearby trajectories in phase space converge or diverge and therefore provide a useful criterion for dynamical stability. In this work, we develop a general framework for analyzing Lyapunov exponents in stationary spacetimes and extend previous studies beyond circular orbits. For a generic stationary spacetime described by $g_{\mu\nu}(r,\theta)$, we show that the phase space relevant to linear perturbations is spanned by $(r,p_r,\theta,p_\theta)$ and the corresponding linear stability matrix has rank four. When the spacetime admits a rank-2 Killing tensor, the linear stability matrix can be reduced to a block-diagonal form, allowing the Lyapunov exponents to be obtained analytically. By imposing the appropriate constraint conditions, the formalism further yields the linear stability matrices and Lyapunov exponents for circular and spherical orbits. The validity of the method is illustrated by several representative examples, including the (inverted) harmonic oscillator, planar Kepler motion, and particle motion in Schwarzschild and Kerr spacetimes. These results provide a unified framework for investigating the stability properties of particle orbits in both classical and relativistic dynamical systems.}


\begin{document} 
	\maketitle

	\section{Introduction} \label{sec:one}
	
The geodesic motion of particles and photons in $(3+1)$-dimensional black-hole spacetimes, with or without a cosmological constant, has been extensively studied. Such geodesics exhibit a rich variety of phenomena and encode important information about the underlying spacetime geometry. For example, the binding energy at the innermost stable circular orbit provides a useful probe of the spin of astrophysical black holes through observations of accretion disks \cite{Narayan:2005ie,Zhang:1997dy,Berti:2015itd}. Likewise, the optical appearance of a star undergoing gravitational collapse depends sensitively on unstable null geodesics \cite{Kokkotas:1999bd,Ianniccari:2024ltb,Konoplya:2017wot}.
	
A particularly important development was reported in Ref.~\cite{Cardoso:2008bp}, in which a direct connection between unstable null geodesics, Lyapunov exponents, and quasinormal modes was established. In particular, the Lyapunov exponent $\lambda$ can be expressed in terms of the radial effective potential as
\begin{align}\label{eq:Lexp}
		\lambda =
		\sqrt{\frac{V_r''(r)}
			{2\,\dot{t}^{\,2}}}.
\end{align}
To derive Eq.~\eqref{eq:Lexp}, one uses the radial equation of motion $\dot{r}^2=V_r$. Here, a prime denotes differentiation with respect to the radial coordinate $r$, while an overdot denotes differentiation with respect to the proper time $\tau$. It is worth emphasizing that Eq.~\eqref{eq:Lexp} applies to equatorial orbits $(\theta=\pi/2)$ in static spherically symmetric and stationary axisymmetric spacetimes.
	
Furthermore, Ref.~\cite{Cardoso:2008bp} demonstrated that, for static, spherically symmetric, asymptotically flat spacetimes, the quasinormal-mode spectrum in the eikonal limit $(l\gg1)$ is determined by the angular velocity $\Omega_c$ of the unstable null geodesic and the Lyapunov exponent \eqref{eq:Lexp}. The corresponding expression is
\begin{align}
		\omega_{\mathrm{QNM}}
		=\Omega_c l
		-i\left(n+\frac{1}{2}\right)|\lambda|.
\end{align}
Here, $n$ denotes the overtone number and $l$ the angular momentum quantum number of the perturbation. It is worth noting that this result follows from the analytic WKB approximation in the eikonal limit \cite{Schutz:1985km,Iyer:1986np,Iyer:1986nq,Ferrari:1984zz,Mashhoon:1985cya,Dolan:2010wr}.
	
For a thermal quantum system, the Lyapunov exponent obeys a universal upper bound $\lambda \le \frac{2 \pi T}{\hbar}$, which scales linearly with the system temperature $T$~\cite{Maldacena:2015waa}. This explicit $T$-dependence was originally derived by analyzing gravitational shock waves propagating near a black hole horizon~\cite{Shenker:2013pqa}. Furthermore, subsequent studies demonstrated that classical particle motion near the horizon likewise exhibits a universal Lyapunov exponent characterizing chaotic dynamics, whose value is governed by the black hole's surface gravity~\cite{Hashimoto:2016dfz,Dalui:2018qqv}.
	
In classical phase space, the Lyapunov exponent measures the asymptotic rate of expansion or contraction of nearby trajectories \cite{layek2015book,dilao2023dynamical}. Specifically, a positive real Lyapunov exponent signifies an exponential divergence between two nearby geodesics, rendering the system's trajectories highly sensitive to changes in initial conditions. In contrast, a purely imaginary Lyapunov exponent indicates that two nearby geodesics remain stable as time evolves. For an extensive overview of Lyapunov exponents in gravitational systems, see Refs.~\cite{Cornish:2003ig,Bombelli:1991eg,Suzuki:1996gm,Levin:1999zx,Chen:2016tmr,Motter:2003jm,Pretorius:2007jn,Perez-Giz:2008ajn,Pradhan:2012rkk,Pradhan:2013bli,Deich:2023oox}.
	
To illustrate these features, we apply Eq.~\eqref{eq:Lexp} to analyze the stability of a circular orbit at $r=r_c$ on the equatorial plane ($\theta = \pi/2$). For an unstable circular orbit for which $V_r''>0$, the Lyapunov exponent $\lambda$ is real, and a linear perturbation evolves as $\delta r = c_0 e^{\lambda t}$. A negative initial perturbation ($c_0 < 0$) causes $r(t)$ to decrease rapidly as the particle plunges into the black hole, whereas a positive perturbation ($c_0 > 0$) drives the orbit to infinity. Conversely, for a stable circular orbit for which with $V_r'' < 0$, $\lambda$ is purely imaginary, yielding a bounded, oscillatory perturbation $\delta r(t) = c_0 e^{i \lambda t}$ with $|\delta r(t)| \le |c_0|$ as $t \to \infty$.
	
However, previous studies leave two major issues unresolved. First, the rank of the linear stability matrix and the dimensionality of the relevant phase space are greater than two. Previous studies typically restricts the analysis to a two-dimensional phase space basis $(r,p_r)$, yielding a $2 \times 2$ matrix that only captures radial perturbations $\delta r$.  However, this subspace represents only part of the full eight-dimensional phase space spanned by $(t,r,\theta,\phi,p_t,p_r,p_\theta,p_\phi)$. 	A complete physical picture requires analyzing the full phase space, which reveals the coupled dynamics of non-radial variations $\delta t$, $\delta \theta$, and $\delta \phi$. Second, previous investigations frequently neglect orbits outside the equatorial plane. The standard formulation holds exclusively for equatorial trajectories, leaving the Lyapunov exponents of non-equatorial trajectories---such as the null spherical orbits essential for characterizing black hole shadows---unexplored.
	
The primary objective of this paper is to resolve these two limitations by providing a comprehensive analysis of the linear stability matrix and Lyapunov exponents in a generic stationary spacetime. To address the phase space dimensionality, we consider a general stationary metric $g_{\mu \nu}(r, \theta)$ and derive the corresponding eight Hamilton equations of motion. Because the metric is independent of $t$ and $\phi$, the corresponding momenta are conserved ($\dot{p}_t=0, \dot{p}_\phi=0$), reducing the perturbation analysis to a six-dimensional basis $(t,r,\theta,\phi,p_r,p_\theta)$. Linearizing the equations of motion reveals that the linear stability matrix $M$ has rank $\mathrm{rank}(M)=4$. This result demonstrates that the minimal phase space governing the perturbation dynamics must be at least four-dimensional. Eliminating the cyclic coordinates $t$ and $\phi$, the physical phase space reduces to $(r,p_r,\theta,p_\theta)$, thereby identifying the relevant degrees of freedom.
	
To address the second issue, we analyze the resulting $4 \times 4$ linear stability matrix within this reduced phase space. Although the full matrix is cumbersome, we render it tractable by introducing a rank-2 Killing tensor to decouple the linear perturbation equations in the $r$ and $\theta$ directions. This decoupling reduces the stability matrix to a block-diagonal form, allowing the Lyapunov exponents to be computed directly. Consequently, we establish a generalized formulation for Lyapunov exponents in stationary spacetimes. Explicit expressions for specific trajectories, such as circular and spherical orbits, can then be obtained directly by imposing the appropriate kinematic constraints on our general result.
	
The linear stability matrix $M_{\rm LP}$ and its corresponding Lyapunov exponents, obtained from Eqs.~\eqref{eq:LPmat} and \eqref{eq:Lya}, take the form
\begin{equation*}
		M_{\rm LP}=
		\renewcommand{\arraystretch}{1.1}
		\left[
		\begin{array}{cc:cc}
			m_{11}  &  m_{12}   & 0 & 0 \\
			m_{21}  & 0 & 0 & 0 \\
			\noalign{\vskip 1mm}
			\hdashline[4pt/3pt]
			0 & 0 & m_{33} &  m_{34}  \\
			0 & 0  & m_{43}   &  0
		\end{array}
		\right],
\end{equation*}
with
\begin{align*}
		\lambda_{1,2}&=\frac{1}{2}\left(m_{11}\mp \sqrt{m_{11}^2+4 m_{12} m_{21}}\right), \\ \lambda_{3,4}&=\frac{1}{2}\left(m_{33}\mp \sqrt{m_{33}^2+ 4 m_{34} m_{43}}\right).
\end{align*}
The explicit matrix elements are shown in Eq.~\eqref{eq:LPele}.
	
The remainder of this paper is organized as follows. In section~\ref{sec:two}, we derive the Hamilton's equations of motion and the linear stability matrix in a generic stationary spacetime to determine the phase space degrees of freedom and derive the generalized Lyapunov exponents.  Section~\ref{sec:three} clarifies the physical interpretation of these exponents using two classical benchmarks: the one-dimensional (inverted) harmonic oscillator and two-dimensional planar Keplerian motion. For both systems, we provide analytical solutions for $r(t)$ and $\delta r(t)$. In sections~\ref{sec:schw} and \ref{sec:kerr}, we extend this framework to relativistic settings, examining timelike and null circular/spherical orbits in Schwarzschild and Kerr black-hole spacetimes. For these cases, we numerically solve for the radial and polar perturbations, $\delta r(\tau)$ and $\delta \theta(\tau)$, and illustrate the trajectories projected onto the $(x,y)$ and $(x,z)$ planes. Finally, section~\ref{sec:conclusion} summarizes our results and discusses their physical implications. Throughout this work, we adopt geometrized units in which $c=\hbar=G=1$.

\section{General stationary black hole spacetimes}\label{sec:two}

A general line element in a stationary spacetime takes the form
\begin{align}
	ds^2=g_{\mu \nu} dx^\mu dx^\nu=g_{tt} dt^2+g_{rr} dr^2+ g_{\theta \theta} d \theta^2+g_{\phi \phi} d \phi^2+ 2 g_{t \phi} dt d \phi.
\end{align}
The metric $g_{\mu \nu}$ is a function of the coordinates $r$ and $\theta$, i.e., $g_{\mu \nu}=g_{\mu \nu}(r,\theta)$. The Lagrangian of a test particle in this spacetime is given by
\begin{align} \label{eq:Lag}
	\mathcal{L} = \frac{1}{2}g_{\mu \nu} \dot{x}^\mu \dot{x}^\nu=\frac{1}{2} \left( g_{tt} \dot{t}^2+g_{rr} \dot{r}^2+ g_{\theta \theta} \dot{\theta}^2+g_{\phi \phi} \dot{\phi}^2+ 2 g_{t \phi} \dot{t} \dot{ \phi} \right).
\end{align}
Using the Lagrangian \eqref{eq:Lag}, the conjugate momentum $p_\mu$ is defined as
\begin{align}\label{conjP}
	p_\mu :=  \frac{ \partial \mathcal{L}}{ \partial \dot{x}^\mu}= g_{ \mu \nu} \dot{x}^\nu.
\end{align}
The corresponding Hamiltonian of the test particle is obtained via the Legendre transformation
\begin{align}\label{eq:Ham}
	\mathcal{H}&=p_\mu \dot{x}^\mu-\mathcal{L}=\frac{1}{2}g^{\mu \nu} p_\mu p_\nu  \nonumber\\
	&= \frac{1}{2}\left( g^{tt} p_t^2+ g^{rr} p_r^2+g^{\theta \theta}p_{\theta}^2+g^{\phi \phi} p_{ \phi }^2+2 g^{t \phi} p_t p_\phi \right).
\end{align}
The inverse metric components appearing in $\mathcal{H}$ follow from $g^{\mu \nu}g_{\nu\alpha}=\delta^\mu_{\;\;\alpha}$
\begin{align*}
	&g^{tt}=\frac{g_{\phi\phi}}{g_{tt} g_{\phi\phi}-g_{t\phi}^2}, \quad g^{\phi \phi}=\frac{g_{tt}}{g_{tt} g_{\phi\phi}-g_{t\phi}^2}, \\
	&g^{t\phi}=\frac{g_{t\phi}}{g_{t\phi}^2-g_{tt} g_{\phi\phi}},\quad g^{rr}=\frac{1}{g_{rr}}, \quad   g^{\theta\theta}=\frac{1}{g_{\theta\theta}} .
\end{align*}

\subsection{Hamilton's equations of motion}

Hamilton's equations of motion can be written as
\begin{align}\label{eq:Heom}
	\dot{x}^\mu = \frac{ \partial \mathcal{H}}{ \partial p_\mu}, \qquad \dot{p}_\mu= -\frac{\partial \mathcal{H}}{ \partial x^\mu}.
\end{align}
Substituting the Hamiltonian \eqref{eq:Ham} into \eqref{eq:Heom} yields the following eight equations
\begin{align}
	\dot{t} &= \frac{\partial \mathcal{H}}{\partial p_t}=f_1(r,\theta):= g^{tt} p_t+g^{t \phi} p_ \phi, \label{eq:td}\\
	\dot{r}&= \frac{\partial \mathcal{H}}{ \partial p_r}=f_2(r,\theta,p_r):=g^{rr} p_r,\\
	\dot{\theta}&= \frac{\partial \mathcal{H}}{ \partial p_\theta}=f_3(r,\theta,p_\theta):=g^{\theta \theta} p_\theta,\\
	\dot{\phi}&= \frac{\partial \mathcal{H}}{ \partial p_\phi}=f_4(r,\theta):=g^{\phi \phi} p_\phi+ g^{t \phi} p_t, \label{eq:phid}\\
	\dot{p}_t&=- \frac{\partial \mathcal{H}}{\partial t}=0,\\
	\dot{p}_r&=- \frac{\partial \mathcal{H}}{\partial r}= f_5 (r,\theta,p_r,p_\theta) \nonumber\\
	&:=\frac{1}{2} \left(-p_\theta^2\, \partial_r g^{\theta \theta} -p_r^2\, \partial_r g^{rr}-p_t^2\, \partial_r g^{tt}-2 p_t p_\phi \,\partial_r g^{t \phi} -p_\phi^2\, \partial_r g^{\phi \phi} \right), \label{eq:prd}\\
	\dot{p}_\theta&=- \frac{\partial \mathcal{H}}{\partial \theta}=f_6( r,\theta,p_r,p_\theta ) \nonumber\\
	&:=\frac{1}{2} \left(-p_\theta^2\, \partial_\theta g^{\theta \theta} -p_r^2\, \partial_\theta g^{rr}-p_t^2\, \partial_\theta g^{tt}-2 p_t p_\phi \,\partial_\theta g^{t \phi} -p_\phi^2\, \partial_\theta g^{\phi \phi} \right), \label{eq:pthd}\\
	\dot{p}_\phi&= -\frac{\partial \mathcal{H}}{\partial \phi}=0.
\end{align}
Here, $f_1,\ldots,f_6$ are regarded as functions of $r$, $\theta$, $p_r$, and $p_\theta$, with the conserved momenta $p_t$ and $p_\phi$ treated as fixed parameters. Note that $\dot{p}_t=0$ and $\dot{p}_\phi=0$ imply that $p_t$ and $p_\phi$ are constants of motion. We identify these conserved quantities as
\begin{align}\label{ELconst}
	p_t=-E, \qquad p_\phi=L.
\end{align}
Thus, the resulting system is described by six nontrivial first-order equations of motion.

\subsection{Linear stability matrix}

To construct the linear stability matrix, we linearize Eqs.~\eqref{eq:td}--\eqref{eq:phid}, \eqref{eq:prd}, and \eqref{eq:pthd}. The coordinates in phase space are $(t,r,\theta,\phi,p_r,p_\theta)$, and we introduce linear perturbations according to
\begin{align}
	(t,r,\theta,\phi,p_r,p_\theta)  \longrightarrow  (t+\delta t,r+ \delta r, \theta+\delta \theta , \phi+\delta \phi, p_r + \delta p_r, p_\theta+ \delta p_\theta).
\end{align}
To first order in the perturbations, the linearized equations are
\begin{align}
	\delta \dot{t} &=  \partial_r f_1\, \delta r+ \partial_\theta f_1\,\delta \theta  ,\\
	\delta \dot{r} &= \partial_r f_2\, \delta  r+\partial_\theta f_2\, \delta  \theta +\partial_{p_r} f_2\, \delta  p_r,\\
	\delta \dot{\theta} &= \partial_r f_3\, \delta  r + \partial_\theta f_3\, \delta  \theta+\partial_{p_\theta} f_3\, \delta  p_\theta,\\
	\delta \dot{\phi} &= \partial_r f_4\,\delta r + \partial_\theta f_4\, \delta  \theta,\\
	\delta \dot{p}_r&= \partial_r f_5\, \delta r+\partial_\theta f_5\, \delta \theta+\partial_{p_r} f_5\, \delta p_r+\partial_{p_\theta} f_5\, \delta p_\theta,\\
	\delta \dot{p}_\theta&= \partial_r f_6\, \delta r+\partial_\theta f_6\, \delta \theta+\partial_{p_r} f_6\, \delta p_r+\partial_{p_\theta} f_6 \delta p_\theta.
\end{align}
The explicit forms of the coefficients are given in appendix~\ref{appA}. The above perturbation equations can be written in matrix form as
\begin{align}\label{eq:mateq}
	\delta \dot{X}=M\,\delta X,
\end{align}
where the perturbation vectors are defined as
\begin{align}
	\delta X &=\left[ \delta t,\delta r, \delta\theta,\delta\phi,\delta p_r, \delta p_\theta\right]^T, \\
 \delta \dot{X}&= \left[ \delta\dot{t},\delta\dot{r}, \delta\dot{\theta}, \delta\dot{\phi}, \delta\dot{p_r}, \delta\dot{p_\theta} \right]^T. 
\end{align}
The matrix $M$ in \eqref{eq:mateq} is the desired linear stability matrix, i.e., the Jacobian matrix, and takes the form
\begin{equation}\label{eq:LSmat}
	M= \frac{\partial (f_1,f_2,f_3,f_4,f_5,f_6) }{\partial (t,r,\theta,\phi,p_r,p_\theta)}=
	\renewcommand{\arraystretch}{1.25}
	\left[
	\begin{array}{
			>{\centering\arraybackslash}m{0.7cm} >{\centering\arraybackslash}m{1cm} >{\centering\arraybackslash}m{1cm} >{\centering\arraybackslash}m{0.8cm} >{\centering\arraybackslash}m{1cm} >{\centering\arraybackslash}m{1cm}
		}
		0 & $\partial_r f_1 $ & $\partial_\theta f_1$ & 0 & 0 & 0 \\
		0 & $\partial_r f_2$ & $ \partial_\theta f_2$ & 0 & $\partial_{p_r} f_2$ & 0 \\
		0 & $ \partial_r f_3 $ & $ \partial_\theta f_3 $& 0 & 0 &  $\partial_{p_\theta} f_3 $ \\
		0 & $\partial_r f_4 $ & $ \partial_\theta f_4 $ & 0 & 0 & 0 \\
		0 & $ \partial_r f_5 $ & $ \partial_\theta f_5 $& 0 & $ \partial_{p_r} f_5 $ & $ \partial_{p_\theta} f_5 $ \\
		0 & $\partial_r f_6$ & $\partial_\theta f_6$ & 0 & $\partial_{p_r} f_6$ & $\partial_{p_\theta} f_6$
	\end{array}
	\right].
\end{equation}
From the linear stability matrix \eqref{eq:LSmat}, we see that all elements in the first and fourth columns vanish because the functions $f_1$ through $f_6$ do not depend on the variables $t$ and $\phi$. Hence the rank of the linear stability matrix $M$ is 4.  This result implies that only four linear perturbation equations are independent. There are two redundant perturbation directions associated with zero eigenvalues.

The reason is that, in a stationary spacetime, $t$ and $\phi$, which appear in the phase-space basis $(t,r,\theta,\phi,p_r,p_\theta)$, are cyclic coordinates. Therefore, their perturbations $\delta t$ and $\delta\phi$ decouple and do not directly affect the dynamical stability. Consequently, the relevant phase space is the subspace spanned by the basis $(r,\theta,p_r,p_\theta)$, which is four-dimensional.

In this new basis $(r,p_r,\theta,p_\theta)$, the linear stability matrix becomes
\begin{equation} \label{eq:subM}
	M= 
	\renewcommand{\arraystretch}{1.3}
	\left[
	\begin{array}{
			>{\centering\arraybackslash}m{1.cm} >{\centering\arraybackslash}m{1.2cm}: >{\centering\arraybackslash}m{1.2cm} >{\centering\arraybackslash}m{1.cm}
		}
		$\partial_r f_2$ & $ \partial_{p_r} f_2 $  &$\partial_\theta f_2$ & 0 \\
		$\partial_r f_5$ & $ \partial_{p_r} f_5 $ &  $\partial_\theta f_5$ & $ \partial_{p_\theta} f_5 $ \\
		\noalign{\vskip 2mm}
		\hdashline[4pt/3pt]
		$ \partial_r f_3$ & 0 & $\partial_\theta f_3 $ &   $\partial_{p_\theta} f_3$ \\
		$\partial_r f_6$ & $\partial_{p_r} f_6$ &  $\partial_\theta f_6$ & $\partial_{p_\theta} f_6 $
	\end{array}
	\right].
\end{equation}
Note that we have changed the ordering of the basis from $(r,\theta,p_r,p_\theta)$ to $(r,p_r,\theta,p_\theta)$, and the resulting matrix also has $\mathrm{rank}(M)=4$. It is worth stressing that this matrix is too complicated to obtain the Lyapunov exponents, i.e., the eigenvalues of the linear stability matrix, in a transparent form. Moreover, because the linear perturbation equations are coupled in the $r$ and $\theta$ variables, this matrix is not block diagonal.

\subsection{Lyapunov exponent}

To simplify the linear stability matrix and evaluate the corresponding
Lyapunov exponents, we focus on the case in which the linear stability
matrix takes a block-diagonal form. This structure can be realized by
introducing a rank-2 Killing tensor in a stationary spacetime. The
existence of such a Killing tensor ensures the separability of the
Hamilton--Jacobi equation. Moreover, the associated additional
conserved quantity, namely the Carter constant, leads to a decoupling
of the equations of motion in the $r$ and $\theta$ directions.

A stationary spacetime admitting a rank-2 Killing tensor can be described by the line element \cite{Hackmann:2011wp,Grenzebach:2014fha}
\begin{align} \label{gLE}
	ds^2=&\frac{-\Delta_r}{\Sigma } \left[ dt-  \chi d \phi \right]^2 +\frac{\sin ^2 \theta \Delta _\theta}{\Sigma } \left[ a dt-\left(r^2+a^2 \right)d \phi \right] ^2   + \frac{\Sigma }{\Delta _r} dr^2 + \frac{\Sigma }{ \Delta_\theta }  d \theta^2,
\end{align}
with 
\begin{align}
	\Sigma  =&r^2+a^2 \cos^2\theta,  \qquad \chi = a \sin ^2 \theta. \label{eq:sigchi}
\end{align}
Here $\Delta_r(r)$ and $\Delta _\theta(\theta)$ are functions of $r$ and $\theta$, respectively. The familiar Kerr--Newman--(A)dS black hole family is recovered by choosing
\begin{align}
	\Delta _\theta = &1+ \frac{\Lambda }{3} a^2 \cos ^2 \theta, \\
	\Delta_r= &r^2-2 m r+ a^2+Q_e^2+Q_m^2 - \frac{\Lambda }{3} \left( a^2 r^2+r^4 \right).
\end{align}
The parameters $m$, $a$, $\Lambda$, $Q_e$, and $Q_m$ denote the black hole mass, angular momentum per unit mass, cosmological constant, electric charge, and magnetic charge, respectively.

In this stationary spacetime, the Killing--Yano tensor takes the form \cite{Kubiznak:2007kh}
\begin{align} \label{eq:KY}
	\textbf{k} =&\frac{1}{2} k_{\mu \nu} \dd x^\mu \wedge \dd x^\nu \nonumber\\
	=& a \cos\theta \dd r \wedge \left[ \dd t -a \sin^2 \theta\dd \phi  \right] - r \sin \theta \dd \theta \wedge \left[a \dd t- (r^2+a^2) \dd \phi  \right].
\end{align}
Contracting the Killing--Yano tensor \eqref{eq:KY} with itself yields the Killing tensor \cite{yano1952some}
\begin{align}\label{eq:KT}
	\mathcal{K}_{\alpha \beta}&= k_{\alpha \gamma } k_\beta{}^\gamma= \begin{bmatrix}
		\mathcal{K}_{tt} & 0 &0 & \mathcal{K}_{t \phi}&\\ 
		0 & \mathcal{K}_{rr} & 0&0 &\\
		0&0 & \mathcal{K}_{\theta \theta} &0 &\\
		\mathcal{K}_{\phi t} & 0 & 0 & \mathcal{K}_{\phi \phi} &\\
	\end{bmatrix}.
\end{align}
Because the nonzero components of $\mathcal{K}_{\alpha\beta}$ are lengthy, we do not display their explicit forms here. Moreover, the Killing tensor is symmetric, i.e., $\mathcal{K}_{\alpha\beta}=\mathcal{K}_{\beta\alpha}$.
One can verify that \eqref{eq:KT} satisfies the Killing tensor equation
\begin{align}
	\nabla_{( \alpha} \mathcal{K}_{\beta \gamma)}=0 .
\end{align}
Using the Killing tensor \eqref{eq:KT}, the Carter constant can be defined as \cite{Carter:1968ks,Carter:1968rr}
\begin{align}
	\mathcal{C}:= \mathcal{K}_{\alpha \beta} \dot{x}^\alpha \dot{x}^\beta.
\end{align}
Since the metric \eqref{gLE} is independent of the coordinates $t$ and $\phi$, two additional conserved quantities arise from the corresponding Killing vectors, which are given by
\begin{align}
	\xi_t^\mu=(1,0,0,0), \qquad  \xi_\phi^\mu=(0,0,0,1).
\end{align}
They give rise to the conserved energy and angular momentum
\begin{align}
	E:= - g_{\mu \nu} \xi_t^\mu \dot{x}^\nu, \qquad L:= g_{\mu \nu} \xi_\phi^\mu  \dot{x}^\nu.
\end{align}
The mass of the test particle is also a conserved quantity,
\begin{align}
	-m_0^2=g_{\mu \nu} \dot{x}^\mu \dot{x}^\nu,
\end{align}
where $m_0=0$ and $m_0=+1$ correspond to the null and timelike cases, respectively. These four constants of motion can be rearranged as 
\begin{align}
	-E=& g_{tt} \dot{t} + g_{t \phi} \dot{\phi}, \label{eq:E} \\
	L=& g_{\phi t} \dot{t}+ g_{\phi \phi} \dot{ \phi}, \label{eq:L}\\
	- m_0^2=& g_{rr} \dot{r}^2+ g_{\theta \theta} \dot{\theta}^2+g_{tt} \dot{t}^2+g_{ \phi \phi} \phi^2+ 2 g_{t \phi} \dot{t} \dot{\phi},\label{eq:m}\\
	\mathcal{C}=& \mathcal{K}_{rr} \dot{r}^2+ \mathcal{K}_{\theta \theta} \dot{\theta}^2+ \mathcal{K}_{tt} \dot{t}^2+ \mathcal{K}_{\phi \phi} \dot{ \phi}^2+2 \mathcal{K}_{t \phi} \dot{t} \dot{\phi} \label{eq:C}.
\end{align}
The first two equations, involving $E$ and $L$, can also be obtained directly from \eqref{conjP} and \eqref{ELconst}. From \eqref{eq:E} and \eqref{eq:L}, we solve for the geodesic equations $\dot{t}$ and $\dot{\phi}$ as
\begin{align}\label{eq:tdphid}
	\dot{t}=\frac{Eg_{\phi\phi}+g_{t \phi} L}{g_{t \phi}^2-g_{tt}g_{\phi\phi}}, \qquad \dot{\phi}= \frac{E g_{t \phi}+g_{tt} L}{g_{tt}g_{\phi\phi}-g_{t \phi}^2}.
\end{align}
The remaining two geodesic equations, $\dot{r}$ and $\dot{\theta}$, can be obtained from \eqref{eq:m} and \eqref{eq:C}:
\begin{align}
	\dot{r}^2=& \frac{\mathcal{C}g_{\theta \theta}-g_{\theta \theta} \left(\mathcal{K}_{tt} \dot{t}^2+2 \mathcal{K}_{t \phi} \dot{t} \dot{\phi}+\mathcal{K}_{\phi \phi} \dot{\phi}^2\right)+\mathcal{K}_{\theta \theta} \left(g_{tt} \dot{t}^2+2 g_{t \phi} \dot{t} \dot{\phi}+g_{\phi \phi } \dot{\phi}^2+m_0 ^2\right)}{g_{\theta \theta} \mathcal{K}_{rr}-g_{rr}K_{\theta \theta}},\label{eq:rd}\\
	\dot{\theta}^2=& \frac{\mathcal{C} g_{rr}-g_{rr}\left(\mathcal{K}_{tt} \dot{t}^2+2 \mathcal{K}_{t \phi} \dot{t} \dot{\phi}+\mathcal{K}_{\phi\phi} \dot{\phi}^2 \right)+ \mathcal{K}_{rr} \left(g_{tt} \dot{t}^2+2 g_{t \phi} \dot{t} \dot{\phi}+g_{\phi \phi}  \dot{\phi}^2+ m_0 ^2 \right)}{g_{rr}\mathcal{K}_{\theta \theta}-g_{\theta \theta} \mathcal{K}_{rr}}. \label{eq:thetad}
\end{align}
For brevity, we do not substitute the expressions in \eqref{eq:tdphid} into the last two equations. Substituting the metric \eqref{gLE} and Killing tensor \eqref{eq:KT} into \eqref{eq:tdphid}, \eqref{eq:rd}, and \eqref{eq:thetad}, the geodesic equations can be written in the following compact form:
\begin{align}
	\Sigma \dot{t} \; &= \frac{r^2+a^2}{\Delta_r}\left( (r^2+a^2)E-a L\right)-\frac{a }{\Delta_\theta }\left(E a \sin ^2 \theta-L\right),\\
	\Sigma  \dot{ \phi} \; &= \frac{a}{\Delta _r} \left( (r^2+a^2)E-a L\right)- \frac{1}{\Delta_\theta \sin ^2 \theta}\left(  E a \sin ^2 \theta-L \right),\\
	\Sigma^2 \dot{r}^2 &= \left( (r^2+a^2)E-aL\right)^2- \Delta_r (m_0^2 r^2+ \mathcal{C}),\\
	\Sigma ^2 \dot{\theta}^2 &= \Delta _\theta \left(\mathcal{C}-m_0^2 a^2 \cos ^2 \theta \right)- \frac{(E a \sin ^2 \theta-L)^2}{ \sin ^2 \theta}.
\end{align}
The right-hand sides of $\Sigma^2\dot{r}^2$ and $\Sigma^2\dot{\theta}^2$ can be defined as
\begin{align} 
	\mathcal{R}(r):=& \left( (r^2+a^2)E-aL\right)^2- \Delta_r (m_0^2 r^2+ \mathcal{C}), \label{eq:Reff} \\
	\Theta(\theta):=&  \Delta _\theta \left(\mathcal{C}-m_0^2 a^2 \cos ^2 \theta \right)- \frac{(E a \sin ^2 \theta-L)^2}{ \sin ^2 \theta} \label{eq:Theff}.
\end{align}
Here, the functions $\mathcal{R}(r)$ and $\Theta(\theta)$ can be regarded as effective potentials for the radial and polar motions, respectively. Therefore, the equations of motion in the $r$ and $\theta$ directions become
\begin{align}
	\left(\frac{d r}{d \lambda} \right)^2 & :=\Sigma ^2 \left( \frac{dr}{d \tau }\right)^2 = \mathcal{R}(r),\\
	\left( \frac{d \theta}{d \lambda } \right)^2 &:=\Sigma ^2 \left( \frac{d \theta}{d \tau }\right)^2 = \Theta(\theta). 
\end{align}
Here, we use the affine parameter $\lambda$ rather than the proper time $\tau$ to absorb the factor $\Sigma=r^2+a^2\cos^2\theta$, which depends on both $r$ and $\theta$. With this new affine parameter, the equations of motion for $\frac{dr}{d\lambda}$ and $\frac{d\theta}{d\lambda}$ are completely decoupled.
The affine parameter $\lambda$ is usually referred to as Mino time \cite{Mino:2003yg}.

Since we focus on the phase space spanned by the basis $(r,p_r,\theta,p_\theta)$, we also require the first-order differential equations for $\frac{dp_r}{d\lambda}$ and $\frac{dp_\theta}{d\lambda}$. Using \eqref{conjP}, one obtains $p_r$ and $p_\theta$ as
\begin{align}
	p_r&=g_{rr} \dot{r}= \frac{\Sigma}{\Delta _r} \dot{r}= \frac{\sqrt{\mathcal{R}(r)}}{\Delta_r}, \label{eq:pr2}\\
	p_\theta&=g_{\theta \theta} \dot{\theta}= \frac{\Sigma }{\Delta _\theta} \dot{\theta}= \frac{\sqrt{\Theta(\theta)}}{\Delta_\theta} \label{eq:pth2}.
\end{align}

\subsubsection{General case}

Differentiating $p_r$ and $p_\theta$ with respect to the Mino time $\lambda$ yields
\begin{align}
	\frac{d p_r}{d \lambda } &= \frac{d }{d r } \left( \frac{\sqrt{\mathcal{R}(r)} }{\Delta_r(r)} \right) \frac{d r}{d \lambda } = \frac{\Delta_r(r) \mathcal{R}'(r)-2 \mathcal{R}(r) \Delta_r'(r)}{2  \Delta_r(r)^2}, \label{eq:pr}\\
	\frac{dp_\theta}{d \lambda } &= \frac{d}{d \theta} \left( \frac{\sqrt{\Theta(\theta)}}{\Delta_\theta(\theta)}\right) \frac{d \theta}{d \lambda } = \frac{\Delta_ \theta (\theta ) \Theta'(\theta )-2 \Theta(\theta ) \Delta_\theta '(\theta )}{2 \Delta_\theta (\theta )^2}.
\end{align}
On the other hand, we can rewrite $\frac{dr}{d\lambda}$ and $\frac{d\theta}{d\lambda}$ as
\begin{align}
	\frac{dr}{d \lambda } =& \Sigma \, \frac{dr }{ d \tau}= \Sigma \times p_r \frac{\Delta_r}{ \Sigma}= p_r \Delta _r, \\
	\frac{d \theta}{ d \lambda } =& \Sigma \, \frac{d \theta}{d \tau}= \Sigma \times p_\theta \frac{\Delta_\theta}{\Sigma}=p_\theta \Delta _\theta.  \label{eq:th}
\end{align}
Using Eqs.~\eqref{eq:pr}--\eqref{eq:th}, the linearized perturbation equations can be obtained as
\begin{equation}\label{eq:LPEQ}
	\begin{aligned}
		\frac{d\; \delta  r}{ d \lambda } &= p_r \Delta'_r(r) \delta r+ \Delta_r \delta p_r,\\
		\frac{d\; \delta  \theta}{ d \lambda } &= p_\theta \Delta_\theta'(\theta) \delta \theta+ \Delta_\theta \delta p_\theta,\\
		\frac{d\; \delta p_r}{d \lambda } &= \frac{d}{dr}\left( \frac{\Delta_r(r) \mathcal{R}'(r)-2 \mathcal{R}(r) \Delta_r'(r)}{2  \Delta_r(r)^2}\right)  \delta r, \\
		\frac{d\; \delta p_\theta}{d \lambda} &= \frac{d}{d \theta} \left(\frac{\Delta_ \theta (\theta ) \Theta'(\theta )-2 \Theta(\theta ) \Delta_\theta '(\theta )}{2 \Delta_\theta (\theta )^2} \right) \delta \theta.
	\end{aligned}
\end{equation}
In the basis $(\delta r,\delta p_r,\delta\theta,\delta p_\theta)$, the linear stability matrix can be read directly from \eqref{eq:LPEQ}
\begin{equation} \label{eq:LPmat}
	M_{LP}=
	\renewcommand{\arraystretch}{1.1}
	\left[
	\begin{array}{
			>{\centering\arraybackslash}m{0.9 cm} >{\centering\arraybackslash}m{0.9cm}: >{\centering\arraybackslash}m{0.9 cm} >{\centering\arraybackslash}m{0.9cm}
		}
		$ m_{11} $ & $ m_{12} $  & 0 & 0 \\
		$ m_{21} $ & 0 & 0 & 0 \\
		\noalign{\vskip 3mm}
		\hdashline[4pt/3pt]
		0 & 0 & $m_{33}$ & $ m_{34} $\\
		0 & 0  & $m_{43}$   &  0
	\end{array}
	\right]
\end{equation}
with
\begin{equation}\label{eq:LPele}
	\begin{aligned}
		& m_{11}=p_r \Delta_r'(r), \quad m_{12}= \Delta _r, \quad m_{33}= p_\theta \Delta '_\theta(\theta), \qquad m_{34}=\Delta _\theta, \\[2mm]
		&m_{21} =\frac{\Delta_r^2 \mathcal{R}''(r)-\Delta_r \left(3 \Delta_r'(r) \mathcal{R}'(r)+2 \mathcal{R}(r) \Delta_r''(r)\right)+4 \mathcal{R}(r) \Delta_r'(r)^2}{2 \Delta_r^3},  \\
		& m_{43} =\frac{\Delta_\theta ^2 \Theta''(\theta )-\Delta_\theta  \left(3 \Delta_\theta '(\theta ) \Theta'(\theta )+2 \Theta(\theta ) \Delta_ \theta ''(\theta )\right)+4 \Theta(\theta ) \Delta_\theta '(\theta )^2}{2 \Delta_\theta^3}. 
	\end{aligned}
\end{equation}
The linear stability matrix \eqref{eq:LPmat} has rank 4, consistent with the result in \eqref{eq:subM}.
Here, however, the matrix takes a block-diagonal form, reflecting the complete decoupling of the linearized perturbation equations in the $r$ and $\theta$ directions.

For the block-diagonal matrix \eqref{eq:LPmat}, the eigenvalues can be obtained straightforwardly. These eigenvalues determine the corresponding Lyapunov exponents, which are given by
\begin{equation}\label{eq:Lya}
	\begin{aligned}
		&\lambda_1=\frac{1}{2}\left(m_{11}- \sqrt{m_{11}^2+4 m_{12} m_{21}}\right),\\
		&\lambda_2=\frac{1}{2}\left(m_{11}+ \sqrt{m_{11}^2+4 m_{12} m_{21}}\right),\\
		&\lambda_3=\frac{1}{2}\left(m_{33}-\sqrt{m_{33}^2+ 4 m_{34} m_{43}}\right),\\
		&\lambda_4=\frac{1}{2}\left(m_{33}+\sqrt{m_{33}^2+ 4 m_{34} m_{43}}\right).
	\end{aligned}
\end{equation}

The linear stability matrix \eqref{eq:LPmat} and the Lyapunov exponents \eqref{eq:Lya} constitute the main results of this paper. They hold for arbitrary trajectories in stationary spacetimes admitting a rank-2 Killing tensor. Specifically, $\lambda_1$ and $\lambda_2$ characterize the linear perturbations in the $(r,p_r)$ sector, while $\lambda_3$ and $\lambda_4$ correspond to those in the $(\theta,p_\theta)$ sector. This pairing reflects the underlying Hamiltonian structure and implies that the phase-space volume is preserved under the linearized evolution.

\subsubsection{Circular orbits}

We first consider the Lyapunov exponents for the simple case of circular orbits.
The corresponding linear stability matrix and Lyapunov exponents can be obtained from the general results \eqref{eq:LPmat} and \eqref{eq:Lya} by imposing the appropriate constraint conditions.

We consider circular orbits located at $r=r_c$ and $\theta=\theta_0$. The radial constraint conditions are
\begin{align}
	\mathcal{R}(r_c)=0, \quad \mathcal{R}'(r_c)=0.
\end{align}
According to \eqref{eq:pr2}, at $r=r_c$ we have
\begin{align}
	p_r= \frac{\sqrt{\mathcal{R}(r_c)}}{\Delta_r(r_c)}=0.
\end{align}
For circular orbits, $\theta$ is constant and does not evolve with time.  Therefore, the conditions for polar motion are
\begin{align}
	\Theta(\theta_0)=0, \quad \Theta'(\theta_0)=0.
\end{align}
Using \eqref{eq:pth2}, we also have
\begin{align}
	p_\theta= \frac{\sqrt{\Theta(\theta_0)}}{\Delta_\theta(\theta_0)}=0.
\end{align}
From the condition $\Theta(\theta_0)=0$ and \eqref{eq:Reff}, one can solve for the Carter constant
\begin{align}
	\Chat= a^2 m_0^2 \cos ^2(\theta_0)+\frac{(L \csc (\theta_0)-a E \sin (\theta_0))^2}{\Delta_\theta (\theta_0)}.
\end{align}
Under the above conditions, the matrix elements of $M_{LP}$ become
\begin{align}
	& m_{11}=0, \qquad m_{12}= \Delta _r(r_c),\qquad m_{21} =\frac{\mathcal{R}''(r_c)}{2 \Delta_r(r_c)}, \\
	& m_{33}= 0, \qquad m_{34}=\Delta _\theta(\theta_0), \qquad  m_{43}= \frac{ \Theta''(\theta_0 )}{2 \Delta_\theta (\theta_0)}.
\end{align}
Using the general formula \eqref{eq:Lya}, the Lyapunov exponents for circular orbits at $r=r_c$ and $\theta=\theta_0$ are
\begin{equation}\label{eq:LyaCO}
	\begin{aligned}
		&\lambda_1=-\sqrt{\frac{\mathcal{R}''(r_c)}{2}},\qquad \lambda_2=+\sqrt{\frac{\mathcal{R}''(r_c)}{2}},\\
		&\lambda_3= - \sqrt{\frac{\Theta''(\theta_0)}{2}}, \qquad \lambda_4= + \sqrt{\frac{\Theta''(\theta_0)}{2}}.
	\end{aligned}
\end{equation}
From Eq.~\eqref{eq:LyaCO}, we see that $\lambda_{1,2}$ and $\lambda_{3,4}$ are governed by the radial and polar effective potentials $\mathcal{R}(r)$ and $\Theta(\theta)$, respectively. For instance, for an unstable circular orbit with $\mathcal{R}''(r_c)>0$, the Lyapunov exponents $\lambda_{1,2}$ are real. By contrast, for a stable circular orbit with $\mathcal{R}''(r_c)<0$, $\lambda_{1,2}$ become purely imaginary. It is worth stressing that the results for $\lambda_1$ and $\lambda_2$ agree with those of previous studies \cite{Cardoso:2008bp,Pradhan:2012rkk}. The exponents $\lambda_{3,4}$ exhibit analogous behavior.

\subsubsection{Spherical orbits}

Next, we consider the Lyapunov exponents for spherical orbits. We consider spherical orbits at a constant radius $r=r_s$, while the polar angle oscillates periodically within $\theta\in[\theta_{\min},\theta_{\max}]$.

In this situation, the radial constraint conditions are
\begin{align}
	\mathcal{R}(r_s)=0, \quad \mathcal{R}'(r_s)=0.
\end{align}
Recall that $p_r=\sqrt{\mathcal{R}(r)}/\Delta_r$, therefore, at $r=r_s$ we also have $p_r=0$. Moreover, when the polar angle $\theta$ reaches its maximum or minimum value, the turning-point condition is
\begin{align}
	\Theta(\theta_{\min})= \Theta(\theta_{\max})=0.
\end{align}
From $\Theta(\theta_{\min})=0$ and \eqref{eq:Theff}, we can immediately solve for the Carter constant
\begin{align}
	\Chat= a^2 m_0^2 \cos ^2(\theta_0)+\frac{(L \csc (\theta_0)-a E \sin (\theta_0))^2}{\Delta_\theta (\theta_0)}.
\end{align}
Here, we set $\theta_0=\theta_{\min}$. After imposing the above conditions, the matrix elements of $M_{LP}$ become
\begin{align}
	& m_{11}=0, \qquad m_{12}= \Delta _r(r_s),\qquad m_{21} =\frac{\mathcal{R}''(r_s)}{2 \Delta_r(r_s)}, \\
	& m_{33}= p_\theta \Delta '_\theta(\theta), \qquad m_{34}=\Delta _\theta,\\
	& m_{43} =\frac{\Delta_\theta^2 \Theta''(\theta )-\Delta_\theta  \left(3 \Delta_\theta '(\theta ) \Theta'(\theta )+2 \Theta(\theta ) \Delta_ \theta ''(\theta )\right)+4 \Theta(\theta ) \Delta_\theta '(\theta )^2}{2 \Delta_\theta ^3}.
\end{align}
Using the general formula \eqref{eq:Lya}, the Lyapunov exponents for spherical orbits with $r=r_s$ and $\theta\in[\theta_{\min},\theta_{\max}]$ are given by
\begin{align}\label{eq:LyaSO}
	&\lambda_1=-\sqrt{\frac{\mathcal{R}''(r_s)}{2}},\\ &\lambda_2=+\sqrt{\frac{\mathcal{R}''(r_s)}{2}},\\
	&\lambda_3=\frac{1}{2}\left(m_{33}-\sqrt{m_{33}^2+ 4 m_{34} m_{43}}\right),\\
	&\lambda_4=\frac{1}{2}\left(m_{33}+\sqrt{m_{33}^2+ 4 m_{34} m_{43}}\right).
\end{align}
The Lyapunov exponents $\lambda_{1,2}$ are determined by the second derivative of the effective potential $\mathcal{R}(r)$. For stable spherical orbits, $\mathcal{R}''(r_s)<0$ implies that $\lambda_{1,2}$ are purely imaginary.  For unstable spherical orbits, $\mathcal{R}''(r_s)>0$, and $\lambda_{1,2}$ become real, implying that the radial perturbation $\delta r$ grows exponentially with Mino time. It is worth emphasizing that the results for $\lambda_{1,2}$ agree with those obtained in \cite{Johnson:2019ljv,Giataganas:2026ayt}. On the other hand, although the expressions for $\lambda_{3,4}$ retain the general form given in \eqref{eq:Lya}, the specific Carter constant $\mathcal{C}=\Chat$ has been imposed.

\section{Classical examples}\label{sec:three}

In this section, we present several classical examples in one and two
dimensions based on the method developed in section~\ref{sec:two}. We
discuss the physical interpretation of the Lyapunov exponent and
examine its relation to the stability of the corresponding orbits.

\subsection{One-dimensional harmonic oscillator}

The first example we consider is the one-dimensional harmonic oscillator, which is a standard system in classical mechanics. The Hamiltonian of the harmonic oscillator is given by
\begin{align} \label{eq:HOham}
	H(p_x,x)=T+V = \frac{p_x^2}{2m}+\frac{1}{2} m \omega^2 x^2.
\end{align}
Using \eqref{eq:HOham}, one obtains Hamilton's equations of motion
\begin{align}
	\frac{d x}{d t} &= \frac{\partial H}{\partial p_x}= \frac{p_x}{m}, \\ \frac{d p_x}{d t}&= - \frac{\partial H}{\partial x}=-m \omega^2 x.
\end{align}
The corresponding linear perturbation equations are
\begin{align}
	\frac{d\, \delta x}{d t} &= \frac{1}{m} \delta p_x, \\
	\frac{d\, \delta p_x}{d t}&=-m \omega^2 \delta  x.
\end{align}
Consequently, the linear stability matrix in the basis $(\delta x,\delta p_x)$ is
\begin{equation}
	M_{HO}=
	\renewcommand{\arraystretch}{1.3}
	\left[
	\begin{array}{
			>{\centering\arraybackslash}m{1 cm} >{\centering\arraybackslash}m{1cm} }
		0 & $ \frac{1}{m} $   \\
		$ -m \omega^2  $ & 0   \\
	\end{array}
	\right].
\end{equation}
The Lyapunov exponents associated with the matrix $M_{HO}$ are
\begin{align}
	\lambda_1=-i \omega, \qquad \lambda _2 =+ i \omega.
\end{align}
From this result, we see that both Lyapunov exponents $\lambda_1$ and $\lambda_2$ are purely imaginary, indicating that the one-dimensional harmonic oscillator is stable under the linear perturbation $\delta x$, as expected.

Let us now examine the trajectories $x_0(t)$ and $x_1(t)=x_0(t)+\delta x(t)$. Here, $x_0(t)$ is the background solution, while $x_1(t)$ is the perturbed trajectory.

From the Euler--Lagrange equation, one obtains
\begin{align}
	\frac{d^2 x}{dt^2}+ \omega^2 x=0.
\end{align}
The general solution to this equation is
\begin{align}
	x_0(t)=A cos(\omega t)+B \sin (\omega t).
\end{align}
Here, $A$ and $B$ are constants. If we consider the perturbed trajectory $x_1(t)=x_0(t)+\delta x(t)$, then $\delta x(t)$ satisfies
\begin{align}
	\frac{d^2 \delta x}{dt^2}+ \omega^2 \delta x=0.
\end{align}
The general solution to this equation is
\begin{align}
	\delta x(t)&= c_1 e^{\lambda _1 t} +c_2 e^{\lambda _2 t} = a \cos (\omega t)+ b \sin (\omega t).
\end{align}
Here, we have used $\lambda_1=-\lambda_2=-i\omega$, together with $a=c_1+c_2$ and $b=i(c_2-c_1)$. It is clear that the solution $\delta x(t)$ is a linear combination of $e^{\lambda_1 t}$ and $e^{\lambda_2 t}$. This shows that the Lyapunov exponents encode the behavior of the linear perturbation.

\begin{figure}[htb]
	\centering
	\includegraphics[width=7cm]{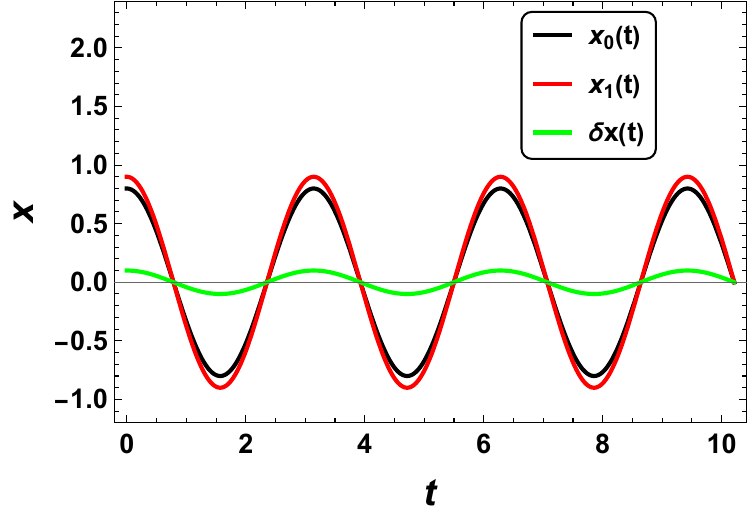}
	\caption{Trajectories of the one-dimensional harmonic oscillator. The curve $x_0(t)$ represents the background solution, while $x_1(t)=x_0(t)+\delta x(t)$ denotes the perturbed trajectory.}
	\label{fig:HO}
\end{figure}

By fixing the parameters $\omega=2$, $A=0.8$, $B=0$, $a=0.1$, and $b=0$, we plot the trajectories $x_0(t)$, $x_1(t)$, and $\delta x(t)$ in figure~\ref{fig:HO}. As can be seen from the figure, the perturbed trajectory remains in the vicinity of the background trajectory over time, indicating that the initial orbit is stable.

\subsection{One-dimensional inverted harmonic oscillator}

We next consider another example: the one-dimensional inverted harmonic oscillator.
The corresponding Hamiltonian is
\begin{align} \label{eq:IHOham}
	H(p_x,x)=T+V = \frac{p_x^2}{2m}-\frac{1}{2} m \omega^2 x^2.
\end{align}
Note that, for the inverted harmonic oscillator, the potential energy $V(r)=-\frac{1}{2} m \omega^2 x^2$ is unbounded from below. This gives rise to the instability of the inverted harmonic oscillator.

Using the Hamiltonian \eqref{eq:IHOham}, one obtains Hamilton's equations of motion
\begin{align}
	\frac{d x}{d t} &= \frac{\partial H}{\partial p_x}= \frac{p_x}{m}, \\
	 \frac{d p_x}{d t}&= - \frac{\partial H}{\partial x}= m \omega^2 x.
\end{align}
The corresponding linear perturbation equations are
\begin{align}
	\frac{d\, \delta x}{d t} &= \frac{1}{m} \delta p_x, \\
	\frac{d\, \delta p_x}{d t}&= m \omega^2 \delta  x.
\end{align}
Consequently, the linear stability matrix in the basis $(\delta x,\delta p_x)$ is
\begin{equation}
	M_{IHO}=
	\renewcommand{\arraystretch}{1.3}
	\left[
	\begin{array}{
			>{\centering\arraybackslash}m{1 cm} >{\centering\arraybackslash}m{1cm} }
		0 & $ \frac{1}{m} $   \\
		$ m \omega^2  $ & 0   \\
	\end{array}
	\right].
\end{equation}
The Lyapunov exponents associated with the matrix $M_{IHO}$ are
\begin{align}
	\lambda_1=-\omega, \qquad \lambda _2 =+ \omega.
\end{align}
This result shows that the Lyapunov exponents of the inverted harmonic oscillator are real, implying that the motion is unstable under the linear perturbation $\delta x$.

Using the Euler--Lagrange equation, we obtain
\begin{align}
	\frac{d^2 x}{dt^2}- \omega^2 x=0.
\end{align}
The general solution to this differential equation is
\begin{align}
	x_0(t)=A e^{\omega t}+B e^ {-\omega t}.
\end{align}
If we consider a perturbed trajectory $x_1(t)=x_0(t)+\delta x(t)$, then $\delta x(t)$ satisfies
\begin{align}
	\frac{d^2 \delta x}{dt^2} - \omega^2 \delta x=0.
\end{align}
The solution to this equation takes the form
\begin{align}
	\delta x(t)&= c_1 e^{\lambda _1 t} +c_2 e^{\lambda _2 t}.
\end{align}
Here, $\lambda_1=-\lambda_2=-\omega$, and the solution $\delta x(t)$ is indeed a linear combination of $e^{\lambda_1 t}$ and $e^{\lambda_2 t}$.

\begin{figure}[htb]
	\centering
	\includegraphics[width=7cm]{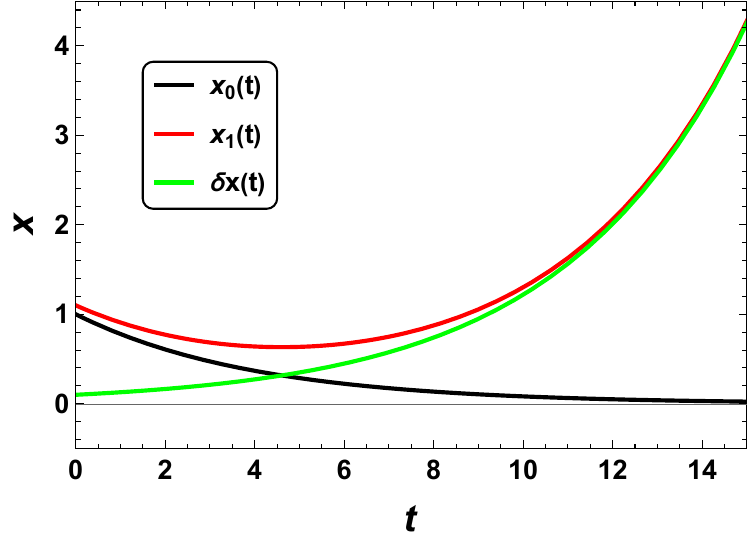}
	\caption{Trajectories of the one-dimensional inverted harmonic oscillator. The curve $x_0(t)$ represents the background solution, while $x_1(t)=x_0(t)+\delta x(t)$ denotes the perturbed trajectory.}
	\label{fig:IHO}
\end{figure}

By fixing the parameters $\omega=1/4$, $A=0$, $B=1$, $c_1=0.1$, and $c_2=0$, we plot the trajectories $x_0(t)$, $x_1(t)$, and $\delta x(t)$ in figure~\ref{fig:IHO}. The background trajectory $x_0(t)$ decays over time, whereas the linearly perturbed trajectory $x_1(t)$ grows rapidly with time and consequently deviates quickly from the initial position.

\subsection{Two-dimensional planar Kepler motion}

The final classical example is a two-dimensional circular orbit in Newtonian gravity. The corresponding Lagrangian is
\begin{align}\label{eq:Klag}
	\mathcal{L}=T-V=\frac{1}{2}m \left( \dot{r}^2+  r^2 \dot{ \phi}^2 \right)+\frac{GMm}{r}.
\end{align}
Here, the gravitational potential is $V(r)=-\frac{GMm}{r}$, and the kinetic energy is expressed in terms of the polar coordinates $(r,\phi)$. The canonical momenta conjugate to $r$ and $\phi$ are
\begin{align}
	p_r= \frac{\partial \mathcal{L}}{\partial \dot{r}}=m \dot{r} \qquad p_\phi=\frac{\partial \mathcal{L}}{\partial \dot{\phi}}=m r^2 \dot{\phi}.
\end{align}
Performing the Legendre transformation, we obtain the Hamiltonian
\begin{align} \label{eq:Kham}
	H= p_r \dot{r}+ p_\phi \dot{\phi}-\mathcal{L}= \frac{p_r^2}{2m}+\frac{p_\phi^2}{2m r^2}- \frac{GMm}{r}.
\end{align}
Using \eqref{eq:Kham}, Hamilton's equations of motion are given by
\begin{align}
	\frac{d r}{dt} &=\quad \frac{\partial H}{\partial p_r}= \frac{p_r}{m}, \\
	\frac{d \phi}{dt} &=\quad \frac{\partial H}{\partial p_\phi}=  \frac{p_\phi}{m r^2} ,  \\
	\frac{d p_r}{dt}&=- \frac{\partial H}{\partial r}=  \frac{p_\phi^2}{m r^3}- \frac{GMm}{r^2},  \\
	\frac{d p_\phi}{dt}&=- \frac{\partial H}{\partial \phi}=   0.
\end{align}
Since $\frac{dp_\phi}{dt}=0$, $p_\phi$ is a conserved quantity, i.e., $p_\phi=L$. The dynamics can be reduced to the phase space spanned by $(r,p_r)$, rather than the full phase space $(r,p_r,\phi,p_\phi)$. The linear stability matrix can then be obtained from the Jacobian matrix
\begin{equation}
	M_{K}= \frac{\partial(F_1,F_2)}{\partial(r,p_r)}=
	\renewcommand{\arraystretch}{1.3}
	\left[
	\begin{array}{
			>{\centering\arraybackslash}m{2.3 cm} >{\centering\arraybackslash}m{1.2cm} }
		0 & $ \frac{1}{m} $   \\
		$ - \frac{3 L^2}{m r^4}+ \frac{2G M m}{r^3}$ & 0  \\
	\end{array}
	\right],
\end{equation}
with
\begin{align}
	F_1(p_r):= \frac{p_r}{m}, \qquad  F_2(r):= \frac{L^2}{m r^3}- \frac{GMm}{r^2} .
\end{align}
The Lyapunov exponents associated with the matrix $M_K$ are
\begin{align}
	\lambda _1 = -\frac{i \sqrt{3 L^2-2 G m^2 M r}}{m r^2}, \qquad \lambda_2 =+\frac{i \sqrt{3 L^2-2 G m^2 M r}}{m r^2}.
\end{align}
For the circular orbit $r=r_c$, the condition $\dot{p}_r=0$ gives the corresponding angular momentum $L^2=GMm^2r_c$. As a result, the Lyapunov exponents for the circular orbit become
\begin{align}
	\lambda _1= -i \sqrt{\frac{G M}{r_c^3}},\qquad \lambda _2=+ i \sqrt{\frac{G M}{r_c^3}}
\end{align}
It is clear that the Lyapunov exponents $\lambda_1$ and $\lambda_2$ associated with the circular orbit $r=r_c$ in Newtonian gravity are purely imaginary. This implies that the circular orbit is stable under linear radial perturbations $\delta r$.

Using the Lagrangian \eqref{eq:Klag}, the radial and angular Euler--Lagrange equations are
\begin{align}
	&m \ddot{r}- m r \dot{\phi}^2+ \frac{GMm}{r^2}=0,\\
	& \frac{d}{dt}\left( m r^2 \dot{\phi} \right)=0,
\end{align}
where the second equations gives the conserved angular momentum $L=mr^2\dot{\phi}$. For the circular orbit $r=r_c$, the solution is
\begin{align}
	r_0(t)&=r_c,\\
	\dot{\phi}^2&=\frac{GM}{r_c^3}.
\end{align}
The solution reads $\phi(t)= \Omega\, t$ with $\Omega=\sqrt{\frac{GM}{r_c^3}}$. If we consider the perturbed trajectory $r_1(t)=r_0(t)+\delta r(t)$ and $\phi_1(t)=\phi(t)+\delta \phi(t)$, then $\delta r(t)$ and $\delta \phi(t)$ satisfy
\begin{align}
	& \delta \ddot{r}+ \Omega^2 \delta r=0,\\
	& \delta \ddot{\phi}+ \frac{2 \Omega}{r_c} \delta \dot{r}=0.
\end{align}
The general solutions to these equations are
\begin{align}
	\delta r(t)  &=c_1 e^{\lambda_1 t}+c_2 e^{\lambda _2 t}\nonumber\\
	&= A \cos  (\Omega t)+ B \sin  (\Omega t),\\
	\delta \phi(t)&= -\frac{2}{r_c} \left[  A\sin (\Omega t)-B \cos (\Omega t) \right]
\end{align}
Here, we have used $\lambda_1=-\lambda_2=-i\Omega$ and defined the constants $A=c_1+c_2$ and $B=i(c_2-c_1)$. The solution $\delta r(t)$ is a linear combination of $e^{\lambda_1 t}$ and $e^{\lambda_2 t}$. Since the Lyapunov exponents are purely imaginary, the circular orbit in Kepler motion is linearly stable.

\begin{figure}[htb]
	\centering
	\includegraphics[width=6.5cm]{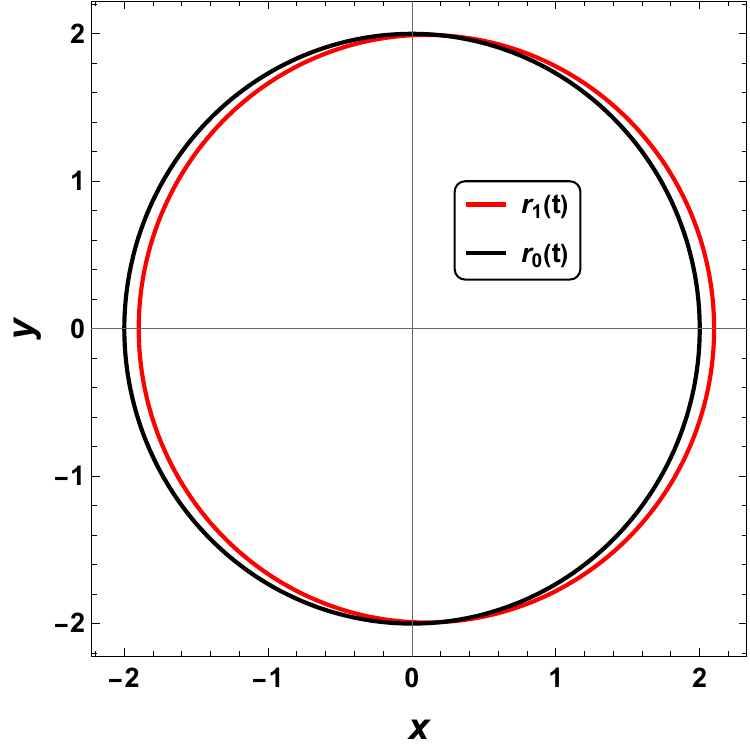}
	\caption{Trajectories of the two-dimensional planar Kepler motion. The curve $r_0(t)$ represents the background solution, while $r_1(t)=r_0(t)+\delta r(t)$ denotes the perturbed trajectory.}
	\label{fig:Kep}
\end{figure}

By choosing the parameters $G=M=1$, $r_c=2$, $A=0.1$, and $B=0$, we plot the trajectories $r_0(t)$ and $r_1(t)$ in figure~\ref{fig:Kep}. The background solution is a circular orbit. After the linear perturbation, the trajectory turns to an ellipse, which indicates the solution is stable.

\section{Relativistic example 1: Schwarzschild black holes} \label{sec:schw}

In section \ref{sec:three}, we investigated several classical examples. We now extend our analysis to relativistic systems, beginning with the Schwarzschild black hole, whose spacetime line element is given by
\begin{align}
	ds^2=&\frac{-\Delta_r}{\Sigma } \left[ dt-  \chi d \phi \right]^2 +\frac{\sin ^2 \theta \Delta _\theta}{\Sigma } \left[ a dt-\left(r^2+a^2 \right)d \phi \right] ^2   + \frac{\Sigma }{\Delta _r} dr^2 + \frac{\Sigma }{ \Delta_\theta }  d \theta^2 \nonumber \\[2mm]
	=&- f  dt^2+ \frac{1}{f} dr^2+ r^2 d \theta^2 + r^2 \sin ^2 \theta d \phi^2, 
\end{align}
with 
\begin{align}
	\Delta_r &=r^2-2m r, \quad \Sigma =r^2, \quad \chi = 0, \quad \Delta_\theta=1,  \quad  f= \frac{\Delta _r}{\Sigma}=1-\frac{2m}{r}.
\end{align}
The effective potentials $\mathcal{R}(r)$ and $\Theta(\theta)$ in Schwarzschild spacetime take the form
\begin{align}
	\mathcal{R}(r)&=E^2 r^4-\left(r^2-2 m r\right) \left(\mathcal{C}+m_0^2 r^2\right), \label{eq:schR}\\
	\Theta(\theta )&=\mathcal{C}-L^2 \csc ^2(\theta ). \label{eq:schTH}
\end{align}
In the basis $(\delta r, \delta p_r,\delta \theta,\delta p_\theta)$, the linear stability matrix from \eqref{eq:LPmat} and \eqref{eq:LPele} is given by

\begin{equation}
	M_{sch}=
	\renewcommand{\arraystretch}{1.1}
	\left[
	\begin{array}{
			>{\centering\arraybackslash}m{0.9 cm} >{\centering\arraybackslash}m{0.9 cm}: >{\centering\arraybackslash}m{0.9 cm} >{\centering\arraybackslash}m{0.9 cm}
		}
		$ m_{11} $ & $ m_{12} $  & 0 & 0 \\
		$ m_{21} $ & 0 & 0 & 0 \\
		
		\noalign{\vskip 3mm}
		\hdashline[4pt/3pt]
		
		0 & 0 & $m_{33}$ & $ m_{34} $\\
		0 & 0  & $m_{43}$   &  0
	\end{array}
	\right],
\end{equation}
with
\begin{align}
	& m_{11}= p_r (2 r-2 m), \qquad m_{12}= r^2-2 mr,\\[2mm]
	&m_{21} =\frac{\mathcal{C} \left(-4 m^3+6 m^2 r-4 m r^2+r^3\right)+2 m^2 r^2 \left(r \left(m_0^2-4 E^2\right)-2 m m_0^2\right)}{r^2 (2 m-r)^3} , \\
	& m_{33}= 0, \qquad m_{34}=1,\\[1mm]
	& m_{43} = -L^2 \csc ^4(\theta )-2 L^2 \cot ^2(\theta ) \csc ^2(\theta ).
\end{align}
The Lyapunov exponents associated with the matrix $M_{sch}$ are
\begin{align}
	&\lambda_1=\frac{1}{2}\left(m_{11}- \sqrt{m_{11}^2+4 m_{12} m_{21}}\right),\\
	&\lambda_2=\frac{1}{2}\left(m_{11}+ \sqrt{m_{11}^2+4 m_{12} m_{21}}\right),\\
	&\lambda_3= -i L\csc ^2(\theta )  \sqrt{\cos (2 \theta )+2}, \\[1mm]
	&\lambda_4= +i L\csc ^2(\theta )  \sqrt{\cos (2 \theta )+2}.
\end{align}
The expressions for $\lambda_1$ and $\lambda_2$ are rather cumbersome, and therefore we do not present them explicitly. Furthermore, we observe that $\lambda_3$ and $\lambda_4$ are purely imaginary. This implies that the polar motion is always linearly stable under perturbations $\delta\theta$. This result is independent of the Carter constant.

\subsection{Lyapunov exponents for circular orbits}

Now let us consider a circular orbit located at $r=r_c$ and $\theta=\theta_0$. The corresponding conditions for the radial and polar motions are
\begin{align}
	\mathcal{R}(r_c)&=0  \quad \mathcal{R}'(r_c)=0,\\
	\Theta(\theta_0)&=0,   \quad \Theta'(\theta_0)=0.
\end{align}
Using the above conditions, we obtain
\begin{align}
	p_r= \frac{\sqrt{\mathcal{R}(r_c)}}{\Delta_r(r_c)}=0, \qquad  p_\theta= \frac{\sqrt{\Theta(\theta_0)}}{\Delta_\theta(\theta_0)}=0.
\end{align}
From $\Theta(\theta_0)=0$ and \eqref{eq:schTH}, the Carter constant can be obtained as
\begin{align}
	\Chat= L^2 \csc^2 (\theta_0).
\end{align}
The condition $\Theta'(\theta_0)=0$ reduces to
\begin{align}
	\Theta'(\theta_0)= 2 L^2 \cot (\theta_0 ) \csc ^2(\theta_0 )=0,
\end{align}
which gives $\theta_0=\frac{\pi}{2}$. This result implies that the circular orbit lies in the equatorial plane $\theta_0=\pi/2$. Under the above constraint conditions, the matrix elements of the linear stability matrix become
\begin{align}
	& m_{11}=0, \qquad m_{12}= \Delta _r(r_c)= r_c^2-2 m r_c,\\
	& m_{21} =\frac{R''(r_c)}{2 \Delta_r(r_c)}=\frac{\Chat+6 r_c \left(r_c \left(m_0^2-E^2\right)-m m_0^2\right)}{r_c (2 m-r_c)},\\
	& m_{33}= 0, \qquad m_{34}=\Delta _\theta( \pi/2 )=1,\\
	& m_{43}= \frac{ \Theta''(\pi/2 )}{2 \Delta_\theta (\pi/2)}=-L^2.
\end{align}
Using the general formula \eqref{eq:LyaCO}, the Lyapunov exponents for a circular orbit with $r=r_c$ and $\theta=\theta_0$ in Schwarzschild spacetime are
\begin{align}
	&\lambda_1=-\sqrt{\frac{\mathcal{R}''(r_c)}{2}}=i \sqrt{6 r_c \left(r_c \left(m_0^2-E^2\right)-m m_0^2\right)+L^2},\\
	&\lambda _2 =+\sqrt{\frac{\mathcal{R}''(r_c)}{2}}=-i \sqrt{6 r_c \left(r_c \left(m_0^2-E^2\right)-m m_0^2\right)+L^2},\\
	&\lambda_3=- \sqrt{\frac{\Theta''( \pi/2)}{2}}=iL,\\
	&\lambda _4= + \sqrt{\frac{\Theta''(\pi/2)}{2}}=-iL.
\end{align}
Note that $E$ and $L$ appearing in the above Lyapunov exponents correspond to the energy and angular momentum of the circular orbit, respectively, and are determined by solving the conditions $\mathcal{R}(r_c)=0$ and $\mathcal{R}'(r_c)=0$. Furthermore, $\lambda_3$ and $\lambda_4$ are purely imaginary, indicating that the circular orbit is linearly stable against $\delta\theta$ perturbations.

\subsubsection{Timelike case}

Having obtained the Lyapunov exponents for circular orbits in Schwarzschild spacetime, we now consider the timelike case with $m_0=1$. The radial effective potential \eqref{eq:schR} becomes
\begin{align}
	\mathcal{R}(r)&=E^2 r^4-\left(r^2-2 m r\right) \left( \Chat+m_0^2 r^2\right) \nonumber\\
	&= E^2 r^4-\left(L^2+r^2\right) \left(r^2-2 m r\right).
\end{align}
From $\mathcal{R}(r_c)=0$ and $\mathcal{R}'(r_c)=0$, the energy and angular momentum at $r=r_c$ can be obtained as
\begin{align}
	E^2= \frac{(r_c-2 m)^2}{r_c (r_c-3 m)},\qquad L^2 = \frac{m\, r_c^2}{r_c-3 m}.
\end{align}
Using the above expressions for $E$ and $L$, and setting $m_0=1$, the Lyapunov exponents for timelike circular orbits take the form
\begin{align}
	\lambda_1 &=-\lambda _2= i \sqrt{\frac{m r_c (r_c-6 m)}{r_c-3 m}},\\[1mm]
	\lambda _3&=-\lambda _4= i r_c \sqrt{\frac{m}{r_c-3 m}}.
\end{align}
For the Schwarzschild black hole, one has $r_c>3m$ and $r_{\mathrm{ISCO}}=6m$. It follows that for $r_c>r_{\mathrm{ISCO}}$, $\lambda_1$ and $\lambda_2$ are purely imaginary. In the range $3m<r_c<r_{\mathrm{ISCO}}$, these two Lyapunov exponents become real. In contrast, $\lambda_3$ and $\lambda_4$ remain purely imaginary in all cases.

\begin{figure}[htb]
	\centering
	\subfigure[]{\includegraphics[width=6cm]{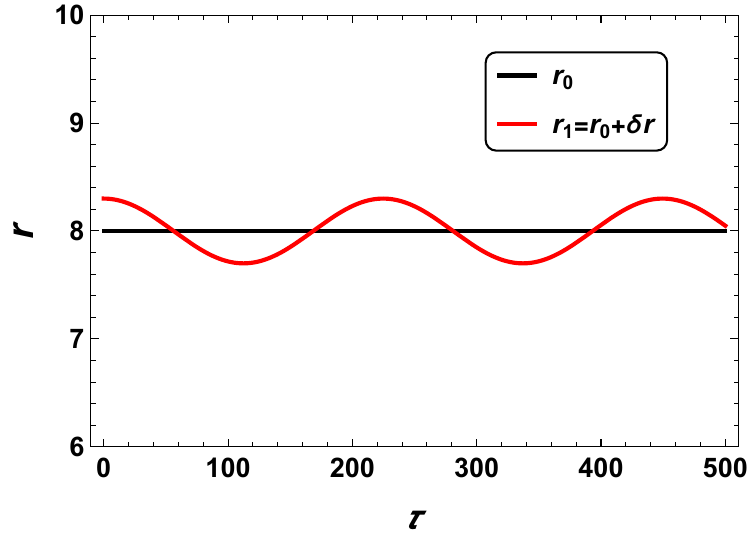}}\hspace{1cm}
	\subfigure[]{\includegraphics[width=6cm]{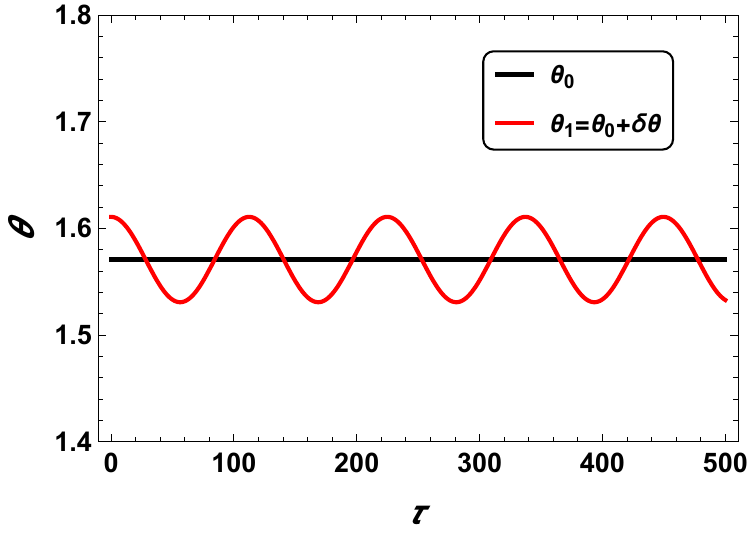}}
	\caption{Timelike circular orbit and its perturbed counterpart in Schwarzschild spacetime. (a) Radial motion with $r_0=8$ and $r_1(\tau)=r_0+\delta r(\tau )$. (b) Polar motion with $\theta_0=\pi/2$ and $\theta_1(\tau )=\theta_0+\delta\theta(\tau )$. The initial perturbations are chosen as $\delta r(0)=0.3$ and $\delta\theta(0)=0.04$.}\label{fig:schTCO}
\end{figure}

To make the above analysis clearer, we present an explicit example by fixing the parameters $m=1$, $r_0=8$, and $\theta_0=\pi/2$. The trajectories $r_1(\tau)$ and $\theta_1(\tau)$, obtained by numerically solving Hamilton's equations of motion, are shown in figure~\ref{fig:schTCO}.

\begin{figure}[htb]
	\centering
	\subfigure{\includegraphics[width=5cm]{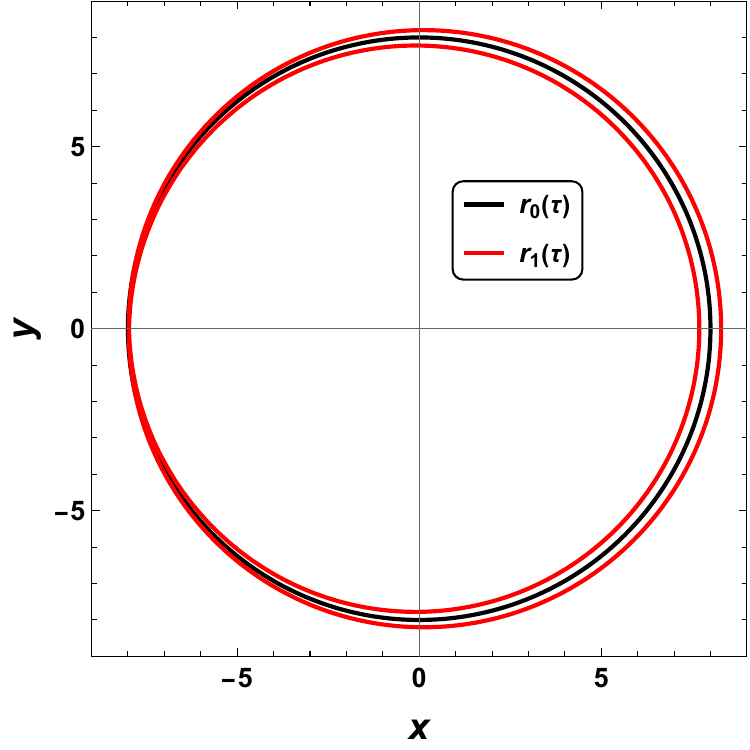}}\hspace{1cm}
	\subfigure{\includegraphics[width=5cm]{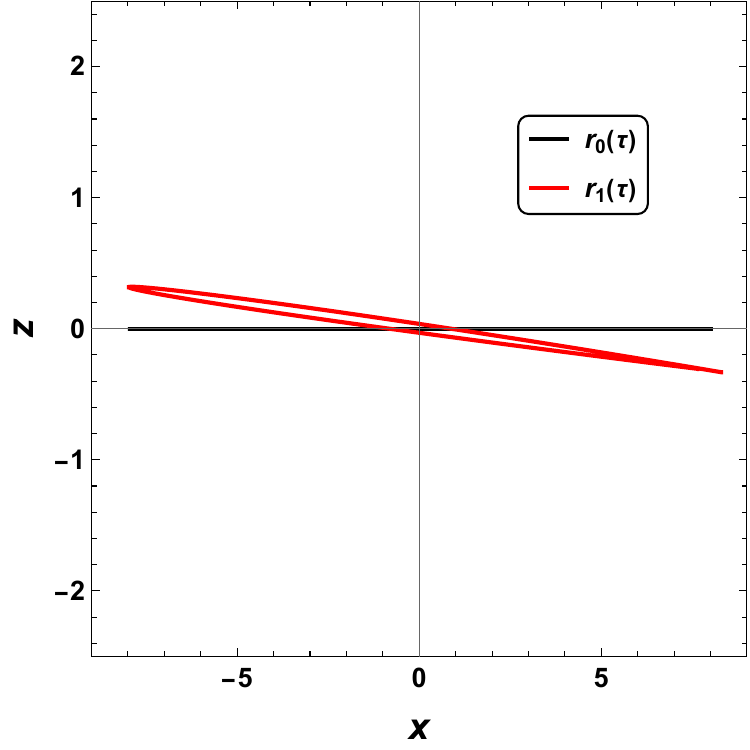}}
	\caption{Projections of a timelike circular orbit and its perturbed orbit onto the $(x,y)$ and $(x,z)$ planes in Schwarzschild spacetime. Left: $(x,y)$ plane. Right: $(x,z)$ plane.} \label{fig:schTxyz}
\end{figure}

The trajectories $r_1 (\tau )$ and $\theta_1 (\tau)$ can be projected onto the $(x,y)$ and $(x,z)$ planes using the following spherical coordinate transformation:
\begin{align}
	x = r \sin\theta \cos\phi, \qquad y = r \sin\theta \sin\phi, \qquad  z = r \cos\theta.
\end{align}
The corresponding projected diagrams are shown in figure~\ref{fig:schTxyz}. Obviously, the circular orbits are stable from the linear perturbation.

\subsubsection{Null case}

For a null circular orbit in Schwarzschild spacetime, we set $m_0=0$. The radial effective potential \eqref{eq:schR} reduces to 
\begin{align}
	\mathcal{R}(r)&=E^2 r^4-\left(r^2-2 m r\right) \left( \Chat+m_0^2 r^2\right) = E^2 r^4-L^2 \left(r^2-2 m r\right).
\end{align}
From the conditions $\mathcal{R}(r_c)=0$ and $\mathcal{R}'(r_c)=0$, we obtain the angular momentum and radius of the null circular orbit
\begin{align}
	L^2= 27 E^2 m^2,\qquad r_c= 3 m.
\end{align}
Note that, for null geodesics, the ratio $L/E$ is invariant under reparametrizations of the affine parameter; therefore, only the relation between $E$ and $L$ is physically relevant. This feature is distinct from the timelike case.

Substituting the above expressions for $L$ and $r_c$ into the Lyapunov exponents, we obtain
\begin{align}
	\lambda _1=-\lambda_2= -3 \sqrt{3} E m,\qquad \lambda _3=-\lambda _4= 3 i \sqrt{3} E m.\label{fafd}
\end{align}
For the null circular orbit at $r_c=3m$ in Schwarzschild spacetime, $\lambda_1$ and $\lambda_2$ are real, indicating instability under radial perturbations $\delta r$. In contrast, $\lambda_3$ and $\lambda_4$ are purely imaginary, implying stability under polar perturbations $\delta \theta$.

\begin{figure}[htb]
	\centering
	\subfigure[]{\includegraphics[width=6cm]{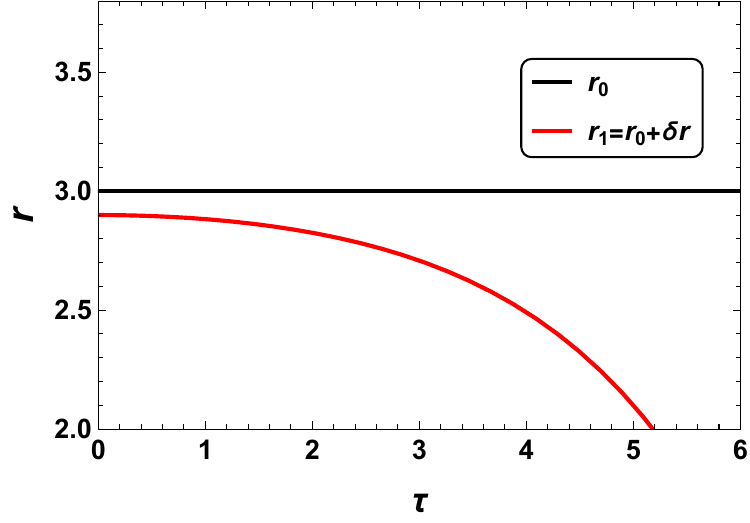}}\hspace{1cm}
	\subfigure[]{\includegraphics[width=6cm]{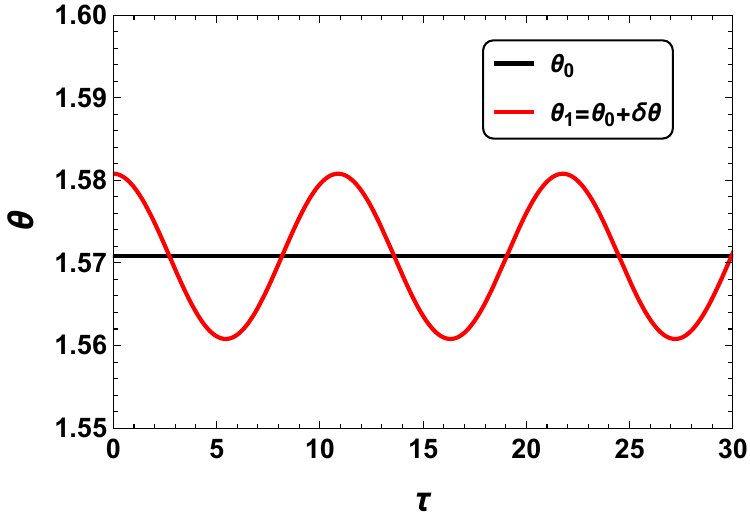}}
	\caption{Null circular orbit and its perturbed counterpart in Schwarzschild spacetime. (a) Radial motion with $r_0=3$ and $r_1(\tau)=r_0+\delta r(\tau)$. (b) Polar motion with $\theta_0=\pi/2$ and $\theta_1(\tau)=\theta_0+\delta\theta(\tau)$. The initial perturbations are chosen as $\delta r(0)=-0.1$ and $\delta\theta(0)=0.01$.} \label{fig:schNCO}
\end{figure}

To illustrate this behavior more clearly, we present a numerical example by fixing $m=1$, $r_0=3$, and $\theta_0=\pi/2$. We solve Hamilton's equations of motion numerically, and the resulting solutions $r_1(\tau)$ and $\theta_1(\tau)$ are shown in figure~\ref{fig:schNCO}.

\begin{figure}[htb]
	\centering
	\subfigure{\includegraphics[width=5cm]{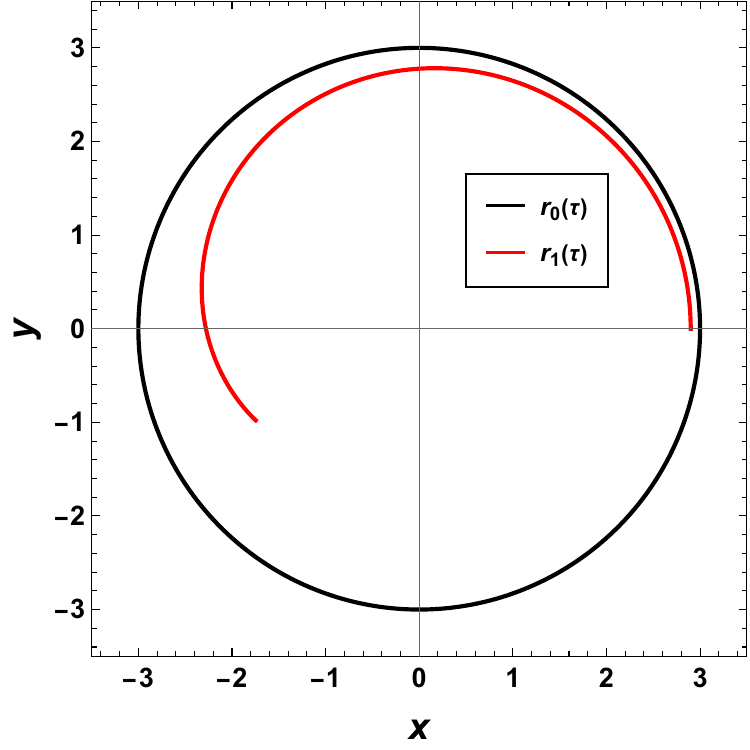}}\hspace{1cm}
	\subfigure{\includegraphics[width=5cm]{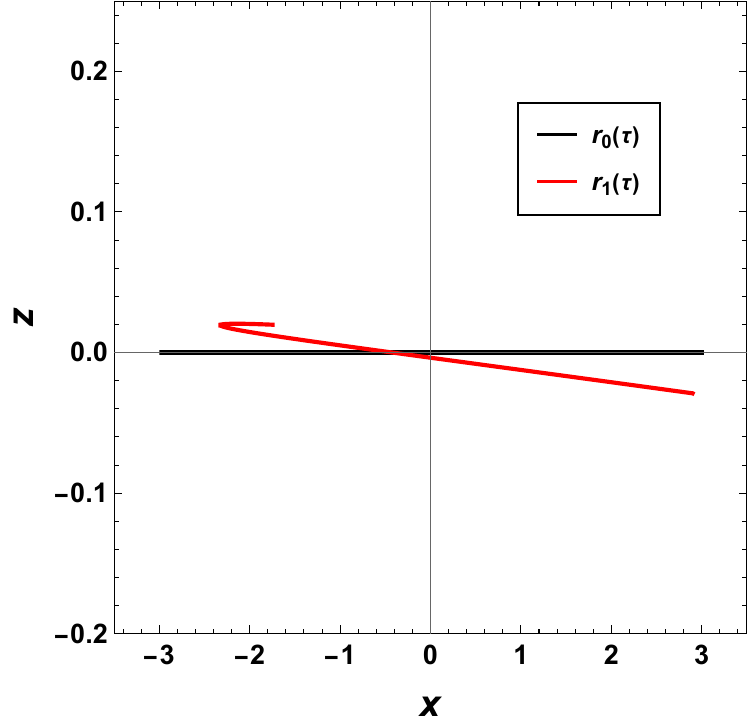}}
	\caption{Projections of a null circular orbit and its perturbed orbit onto the $(x,y)$ and $(x,z)$ planes in Schwarzschild spacetime. Left: $(x,y)$ plane. Right: $(x,z)$ plane.} 
	\label{fig:schNxyz}
\end{figure}

The projections of the trajectories $r_1(\tau)$ and $\theta_1(\tau)$ onto the $(x,y)$ and $(x,z)$ planes are shown in figure~\ref{fig:schNxyz}. It is clear that the radial motion is unstable, while the $\theta$ motion is stable, which is consistent with teh Lyapunov exponents obtained in Eq. (\ref{fafd}).

\subsection{Lyapunov exponents for spherical orbits}

We assume that the spherical orbits are located at $r=r_s$, while $\theta$ oscillates between $\theta_{\min}$ and $\theta_{\max}$. The radial conditions for spherical orbits are
\begin{align}
	\mathcal{R}(r_s)=0, \quad \mathcal{R}'(r_s)=0.
\end{align}
The first condition implies that $p_r=\sqrt{\mathcal{R}(r_s)}/\Delta_r(r_s)=0$ at $r=r_s$. Moreover, at the turning points $\theta_{\min}$ and $\theta_{\max}$, one has $\Theta(\theta_{\min})= \Theta(\theta_{\max})=0$. Then the Carter constant is obtained as
\begin{align}
	\Chat = L^2 \csc^2 (\theta_0),
\end{align}
where we denote $\theta_0=\theta_{\min}$. Upon imposing the above conditions, the matrix elements of the linear stability matrix become
\begin{align}
	& m_{11}=0, \qquad m_{12}= \Delta _r(r_s)=r_s^2-2m r_s,\\[1mm]
	& m_{21} =\frac{R''(r_s)}{2 \Delta_r(r_s)}=\frac{\Chat+6 r_s \left(r_s \left(m_0^2-E^2\right)-m m_0^2\right)}{r_s (2 m-r_s)}, \\
	& m_{33}= 0 , \qquad m_{34}=1,\\[1mm]
	& m_{43} = -L^2 \csc ^4(\theta )-2 L^2 \cot ^2(\theta ) \csc ^2(\theta ).
\end{align}
From the general formula \eqref{eq:LyaSO}, the Lyapunov exponents for a spherical orbit with $r=r_s$ and $\theta\in[\theta_{\min},\theta_{\max}]$ are
\begin{align}
	\lambda _1&=-\sqrt{\mathcal{R}''(r_s)/2} = i \sqrt{6 r_s \left(r_s \left(m_0^2-E^2\right)-m m_0^2\right)+L^2 \csc ^2(\theta )},\\[1mm]
	\lambda _2 &=+\sqrt{\mathcal{R}''(r_s)/2}= -i \sqrt{6 r_s \left(r_s \left(m_0^2-E^2\right)-m m_0^2\right)+L^2 \csc ^2(\theta )},\\[1mm]
	\lambda _3 &=i L \sqrt{\cos (2 \theta )+2} \csc ^2(\theta ),\\[1mm]
	\lambda _4 &= -i L \sqrt{\cos (2 \theta )+2} \csc ^2(\theta ).
\end{align}
Note that we have used $\mathcal{C}=\hat{\mathcal{C}}$. The quantities $E$ and $L$ appearing in the Lyapunov exponents are the energy and angular momentum of the spherical orbit, respectively, and can be evaluated from $\mathcal{R}(r_s)=0$ and $\mathcal{R}'(r_s)=0$.

\subsubsection{Timelike case}

Let us consider timelike spherical orbits by setting $m_0=1$. The radial effective potential from \eqref{eq:schR} becomes
\begin{align}
	\mathcal{R}(r)&=E^2 r^4-\left(r^2-2 m r\right) \left( \Chat+m_0^2 r^2\right) \nonumber\\
	&= E^2 r^4-\left(r^2-2 m r\right) \left(L^2 \csc ^2(\theta_0 )+r^2\right) .
\end{align}
From $\mathcal{R}(r_s)=0$ and $\mathcal{R}'(r_s)=0$, the energy and angular momentum can be obtained as
\begin{align}
	E^2= \frac{(r_s-2 m)^2}{r_s (r_s-3 m)}, \qquad L^2= \frac{m r_s^2 \sin ^2(\theta_0 )}{r_s-3 m}.
\end{align}
Using the above expressions for $E$ and $L$, the Lyapunov exponents for timelike spherical orbits are
\begin{align}
	& \lambda _1 = -\lambda _2= i \sqrt{\frac{m r_s (6 m-r_s)}{3 m-r_s}}, \\
	& \lambda _3=-\lambda _4= i r_s \csc ^2(\theta ) \sin (\theta_0) \sqrt{\frac{m (\cos 2 \theta +2)}{r_s-3 m}}.
\end{align}
For spherical orbits in Schwarzschild spacetime, the Lyapunov exponents $\lambda_1$ and $\lambda_2$ are the same as those in the timelike circular orbit case. If $r_s>6m$, $\lambda_1$ and $\lambda_2$ are purely imaginary. If $3m<r_s<6m$, the exponents become real. However, for spherical orbits, $\lambda_3$ and $\lambda_4$ depend on both the turning point $\theta_0$ and the polar angle $\theta$.

\begin{figure}[htb]
	\centering
	\subfigure[]{\includegraphics[width=6cm]{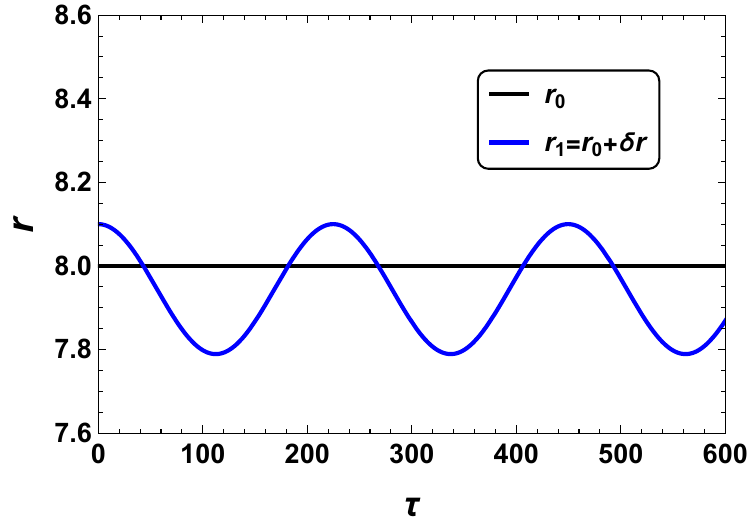}}\hspace{1cm}
	\subfigure[]{\includegraphics[width=6cm]{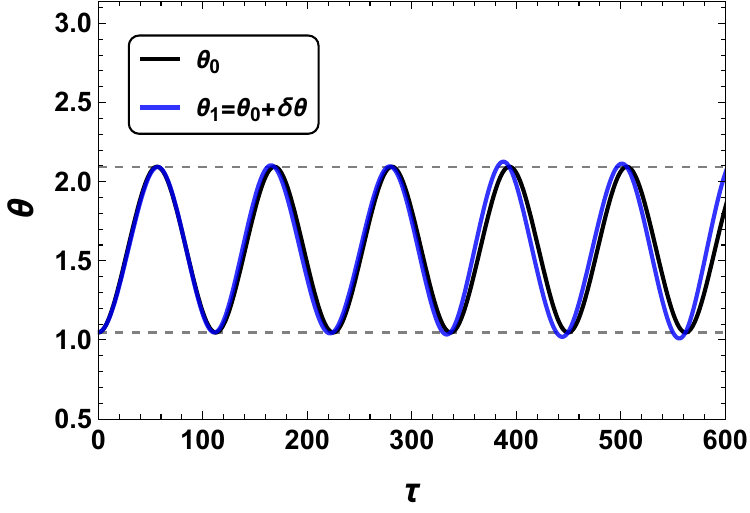}}
	\caption{Timelike spherical orbit and its perturbed counterpart in Schwarzschild spacetime. (a) Radial motion with $r_0=8$ and $r_1(\tau)=r_0+\delta r(\tau)$. (b) Polar motion with $\theta (\tau)\in [ \frac{\pi}{3},\frac{2 \pi}{3}]$ and $\theta_1(\tau)=\theta (\tau)+\delta\theta(\tau )$. The initial perturbations are chosen as $\delta r(0)=0.1$ and $\delta\theta(0)=0.002$.}
	\label{fig:schTSO}
\end{figure}

\begin{figure}[htb]
	\centering
	\subfigure{\includegraphics[width=5cm]{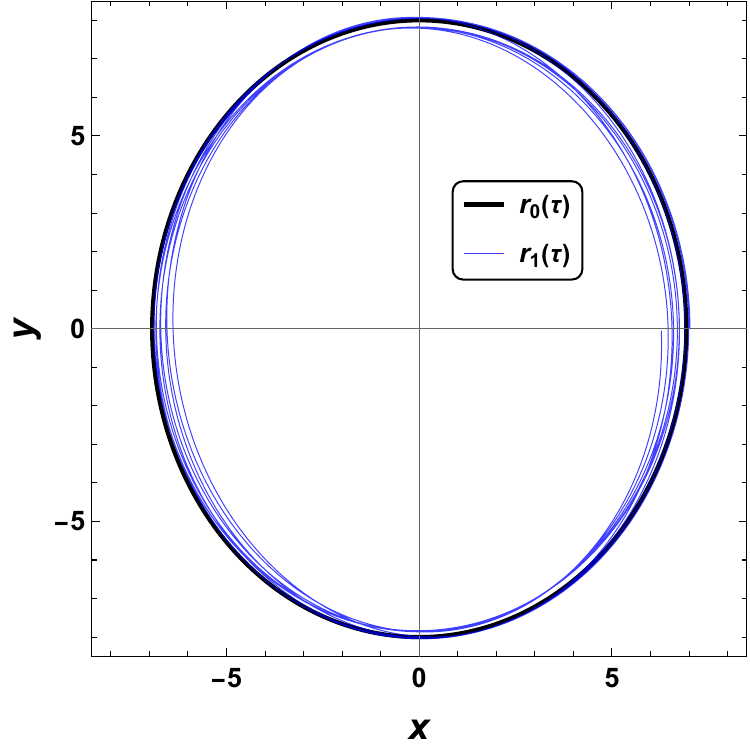}}\hspace{1cm}
	\subfigure{\includegraphics[width=5cm]{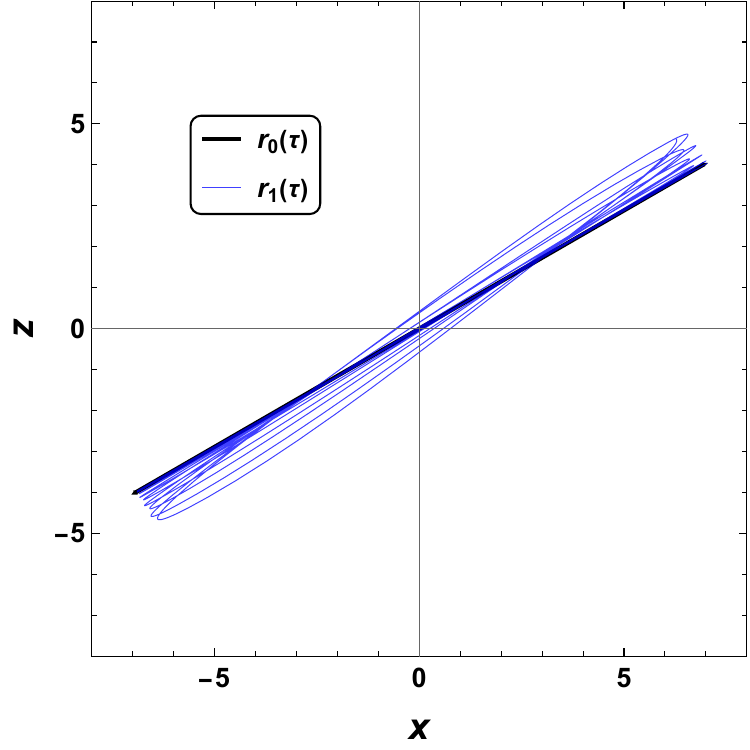}}
	\caption{Projections of a timelike spherical orbit and its perturbed orbit onto the $(x,y)$ and $(x,z)$ planes in Schwarzschild spacetime. Left: $(x,y)$ plane. Right: $(x,z)$ plane.} 
	\label{fig:schTSxyz}
\end{figure}

We present an explicit example by fixing the parameters $m=1$, $r_0=8$, and $\theta_0=\pi/3$. The polar angle oscillates in the range $\theta\in[ \frac{\pi}{3}, \frac{2 \pi}{3}]$. By numerically solving Hamilton's equations of motion, we obtain the trajectories $r_1(\tau)$ and $\theta_1(\tau)$ shown in figure~\ref{fig:schTSO}. The projections of the trajectories $r_1(\tau)$ and $\theta_1(\tau)$ onto the $(x,y)$ and $(x,z)$ planes are shown in figure~\ref{fig:schTSxyz}, from which, one can find that both the motions are stable.

\subsubsection{Null case}

Let us consider a null spherical orbit in Schwarzschild spacetime by setting $m_0=0$. The radial effective potential from \eqref{eq:schR} takes the form
\begin{align}
	\mathcal{R}(r) =E^2 r^4-\left(r^2-2 m r\right) \left( \Chat+m_0^2 r^2\right) = E^2 r^4-L^2 \csc ^2(\theta_0) \left(r^2-2 m r\right) .
\end{align}
From the conditions $\mathcal{R}(r_s)=0$ and $\mathcal{R}'(r_s)=0$, one can obtain the angular momentum and radius of the null spherical orbit
\begin{align}
	L^2= 27 E^2 m^2 \sin ^2(\theta_0),\qquad  r_s= 3 m.
\end{align}
Note that, in the null case, the ratio $L/E$ is invariant under reparametrizations of the affine parameter.

Using the above expressions for the angular momentum $L$ and orbital radius $r_s$, we obtain the Lyapunov exponents for null spherical orbits
\begin{align}
	&\lambda _1=-\lambda _2 =-3 \sqrt{3} E m, \label{LyaschNSOr} \\
	&\lambda _3= -\lambda _4=  3 i \sqrt{3} E m \sqrt{\cos (2 \theta )+2} \csc ^2(\theta ) \sin (\theta_0). \label{LyaschNSOth}
\end{align}
For the null spherical orbit at $r_s=3m$ in Schwarzschild spacetime, $\lambda_1$ and $\lambda_2$ are real, indicating that the orbit is unstable under radial perturbations $\delta r$. In contrast, $\lambda_3$ and $\lambda_4$ are purely imaginary, implying stability under perturbations $\delta \theta$.

\begin{figure}[htb]
	\centering
	\subfigure[]{\includegraphics[width=6cm]{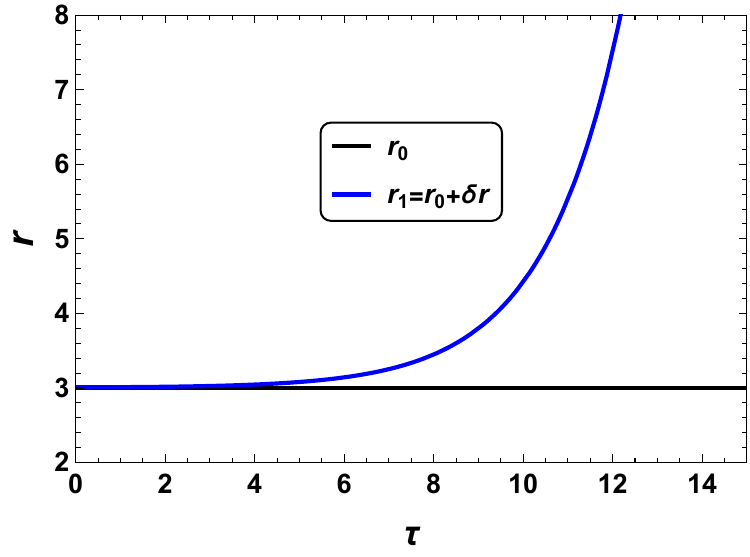}}\hspace{1cm}
	\subfigure[]{\includegraphics[width=6cm]{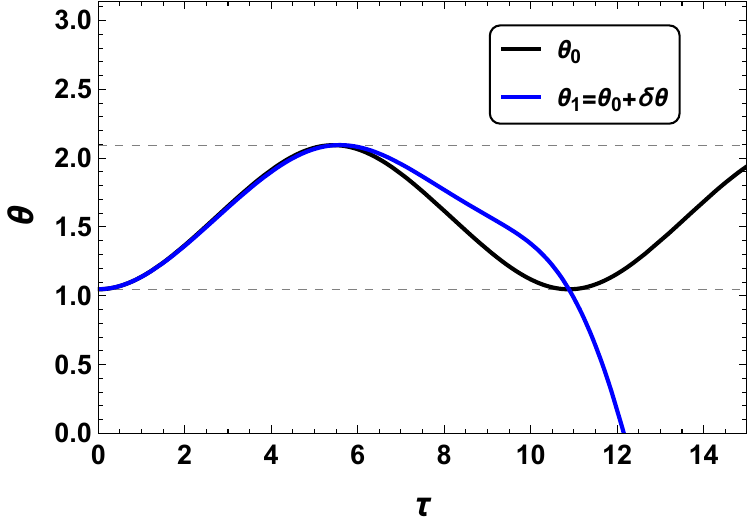}}
	\caption{Null spherical orbit and its perturbed counterpart in Schwarzschild spacetime. (a) Radial motion with $r_0=3$ and $r_1(\tau )=r_0+\delta r(\tau )$. (b) Polar motion with $\theta(\tau ) \in [ \frac{\pi}{3},\frac{2 \pi}{3}]$ and $\theta_1(\tau )=\theta (\tau )+\delta\theta(\tau)$. The initial perturbations are chosen as $\delta r(0)=0.01$ and $\delta\theta(0)=0.001$.}
	\label{fig:schNSO}
\end{figure}

\begin{figure}[htb]
	\centering
	\subfigure{\includegraphics[width=5cm]{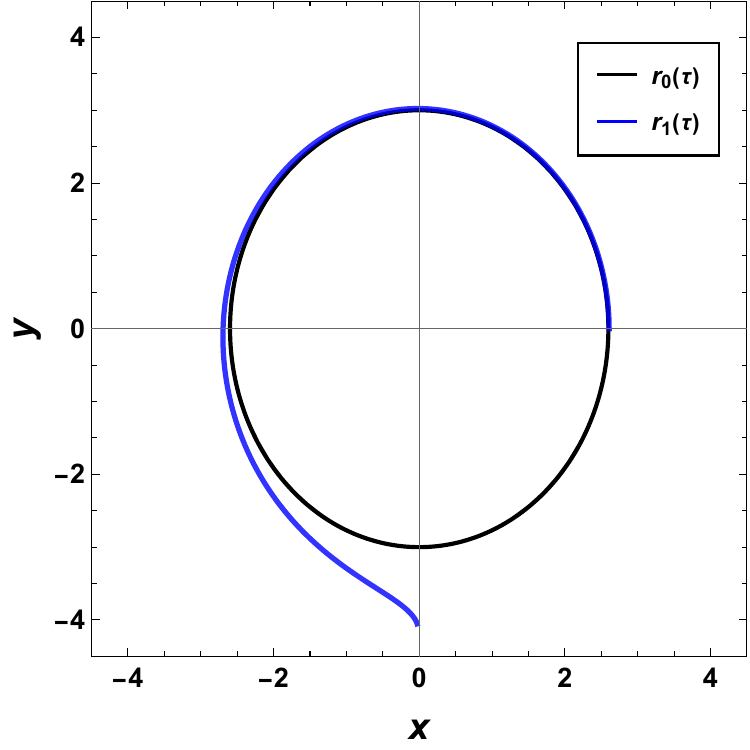}}\hspace{1cm}
	\subfigure{\includegraphics[width=5cm]{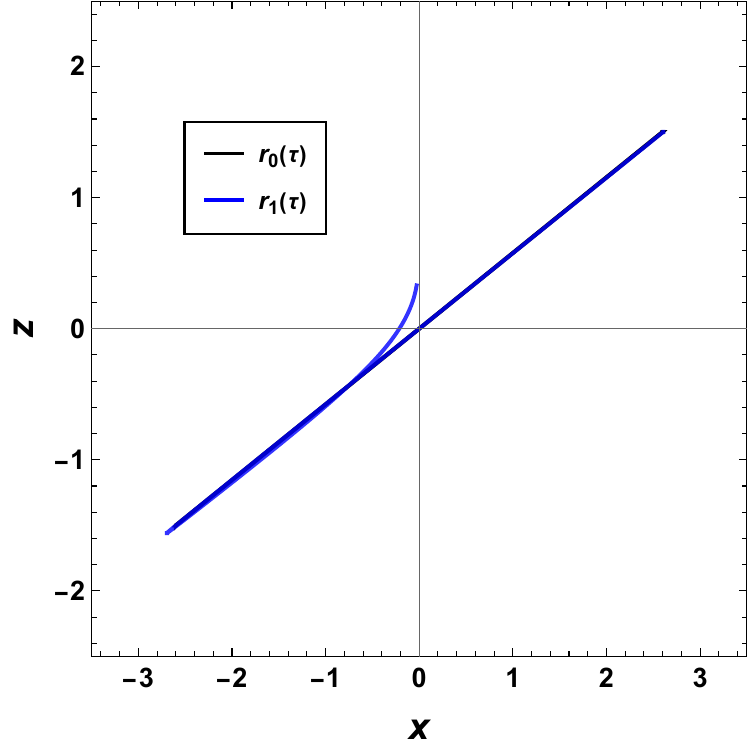}}
	\caption{Projections of a timelike spherical orbit and its perturbed orbit onto the $(x,y)$ and $(x,z)$ planes in Schwarzschild spacetime. Left: $(x,y)$ plane. Right: $(x,z)$ plane.} 
	\label{fig:schNSxyz}
\end{figure}

As an explicit example, we choose the parameters $m=1$, $r_0=3$, and $\theta_0=\pi/3$, with the polar angle allowed to vary within the range $[\pi/3, 2\pi/3]$. The time evolution of the radial trajectory $r_1(\tau)$ and the polar trajectory $\theta_1(\tau)$ is shown in figure \ref{fig:schNSO}. The projections of these trajectories onto the $(x,y)$ and $(x,z)$ planes are shown in figure \ref{fig:schNSxyz}.

As shown in figure \ref{fig:schNSO}, the perturbed radial trajectory $r_1(\tau)$ rapidly diverges from the background solution $r_0(\tau)$. This indicates that the radial perturbation is unstable, consistent with the Lyapunov exponents $\lambda_{1,2}$ in Eq.~\eqref{LyaschNSOr}. For the polar motion, $\theta_1(\tau)$ remains close to $\theta_0(\tau)$ at early times ($\tau < 6$), indicating that the polar perturbation is initially stable, in agreement with the Lyapunov exponents $\lambda_{3,4}$ in Eq.~\eqref{LyaschNSOth}. However, at later times ($\tau > 6$), $\theta_1(\tau)$ begins to deviate significantly from $\theta_0(\tau)$. This behavior can be attributed to the growth of the radial perturbation, for which $\delta r = |r_1-r_0| \sim r_0$, thereby violating the perturbative condition $\delta r \ll r_0$. Consequently, the linear perturbation analysis is no longer valid at sufficiently late times.

\section{Relativistic example 2: Kerr black holes} \label{sec:kerr} 

In the previous section \ref{sec:schw}, we discussed the Schwarzschild black hole, which describes a static and spherically symmetric spacetime. In this section, we turn our attention to the Kerr black hole, characterized by two parameters: mass $m$ and spin $a$. The Kerr spacetime is stationary and axisymmetric, and the corresponding line element is given by
\begin{align}\label{eq:kerrLE}
	ds^2=&\frac{-\Delta_r}{\Sigma } \left[ dt-  \chi d \phi \right]^2 +\frac{\sin ^2 \theta \Delta _\theta}{\Sigma } \left[ a dt-\left(r^2+a^2 \right)d \phi \right] ^2   + \frac{\Sigma }{\Delta _r} dr^2 + \frac{\Sigma }{ \Delta_\theta }  d \theta^2 ,
\end{align}
with 
\begin{align}
	&\Sigma =r^2+a^2 \cos ^2 \theta, \quad \chi = a \sin ^2 \theta,\\
	&\Delta_r =r^2-2m r+a^2,  \quad  \Delta_\theta=1 .
\end{align}
Using the Kerr metric \eqref{eq:kerrLE}, the effective potentials $\mathcal{R}(r)$ and $\Theta(\theta)$ from \eqref{eq:Reff} and \eqref{eq:Theff} are given by
\begin{align}
	\mathcal{R}(r)&=\left(E \left(a^2+r^2\right)-a L\right)^2-\left(a^2-2 m r+r^2\right) \left(\mathcal{C}+m_0^2 r^2\right), \label{eq:kerrR}\\
	\Theta(\theta )&=\mathcal{C}-a^2 m_0^2 \cos ^2(\theta )-\csc ^2(\theta ) \left(L-a E \sin ^2\theta \right)^2. \label{eq:kerrTH}
\end{align}
In the basis $(\delta r,\delta p_r,\delta \theta,\delta p_\theta)$, the linear stability matrix from \eqref{eq:LPmat} and \eqref{eq:LPele} takes the form 
\begin{equation}
	M_{Kerr}=
	\renewcommand{\arraystretch}{1.1}
	\left[
	\begin{array}{
			>{\centering\arraybackslash}m{0.9 cm} >{\centering\arraybackslash}m{0.9cm}: >{\centering\arraybackslash}m{0.9 cm} >{\centering\arraybackslash}m{0.9 cm}
		}
		$ m_{11} $ & $ m_{12} $  & 0 & 0 \\
		$ m_{21} $ & 0 & 0 & 0 \\
		\noalign{\vskip 3mm}
		\hdashline[4pt/3pt]
		0 & 0 & $m_{33}$ & $ m_{34} $\\
		0 & 0  & $m_{43}$   &  0
	\end{array}
	\right],
\end{equation}
with
\begin{align}
	& m_{11}= p_r (2 r-2 m), \qquad m_{12}= r^2+a^2-2 mr,\\[2mm]
	&m_{21} =\frac{\Delta_r^2 \mathcal{R}''(r)-\Delta_r \left(3 \Delta_r'(r) \mathcal{R}'(r)+2 \mathcal{R}(r) \Delta_r''(r)\right)+4 \mathcal{R}(r) \Delta_r'(r)^2}{2 \Delta_r^3}, \\
	& m_{33}= 0, \qquad m_{34}=1,\\
	& m_{43} =\frac{\Delta_\theta ^2 \Theta''(\theta )-\Delta_\theta  \left(3 \Delta_\theta '(\theta ) \Theta'(\theta )+2 \Theta(\theta ) \Delta_ \theta ''(\theta )\right)+4 \Theta(\theta ) \Delta_\theta '(\theta )^2}{2 \Delta_\theta^3}.
\end{align}
Here, we do not present the explicit expressions for $m_{21}$ and $m_{43}$ because of their length. From the above matrix $M_{Kerr}$, we obtain the Lyapunov exponents
\begin{align}
	&\lambda_1=\frac{1}{2}\left(m_{11}- \sqrt{m_{11}^2+4 m_{12} m_{21}}\right),\\
	&\lambda_2=\frac{1}{2}\left(m_{11}+ \sqrt{m_{11}^2+4 m_{12} m_{21}}\right),\\
	&\lambda_3=-\sqrt{m_{43}}, \quad \qquad  \lambda_4=+ \sqrt{m_{43}} .
\end{align}
In contrast to $\lambda_1$ and $\lambda_2$, the Lyapunov exponents
$\lambda_3$ and $\lambda_4$ take a particularly simple form owing to $m_{33}=0$ and $m_{34}=1$.

\subsection{Lyapunov exponents for circular orbits}

We consider a circular orbit located at $r=r_c$ and $\theta=\theta_0$ in the Kerr spacetime. The conditions for radial and polar motions are given by
\begin{align}
	\mathcal{R}(r_c)&=0,  \quad \mathcal{R}'(r_c)=0,\\
	\Theta(\theta_0)&=0,   \quad \Theta'(\theta_0)=0.
\end{align}
Using the above conditions, it follows that
\begin{align}
	p_r= \frac{\sqrt{\mathcal{R}(r_c)}}{\Delta_r(r_c)}=0, \qquad  p_\theta= \frac{\sqrt{\Theta(\theta_0)}}{\Delta_\theta(\theta_0)}=0.
\end{align}
From $\Theta(\theta_0)=0$ and \eqref{eq:kerrTH}, the Carter constant can be obtained as
\begin{align}
	\Chat= a^2 m_0^2 \cos ^2\theta_0+(L \csc \theta_0-a E \sin \theta_0)^2.
\end{align}
The value of $\theta_0$ can be determined from $\Theta'(\theta_0)=0$
\begin{align}
	\theta_0=\frac{\pi}{2}.
\end{align}
This result implies that the circular orbit is confined to the
equatorial plane with $\theta_0=\pi/2$. With the above conditions, the matrix elements of $M_{Kerr}$ reduce to
\begin{align}
	& m_{11}=0, \qquad m_{12}= \Delta _r(r_c)= r_c^2-2 m r_c+a^2,\\
	& m_{21} =\frac{\mathcal{R}''(r_c)}{2 \Delta_r(r_c)}= \frac{a^2 \left(2 E^2-m_0^2\right)-2 a E L-\Chat+6 r_c \left(E^2 r_c+m m_0^2-m_0^2 r_c\right)}{a^2+r_c (r_c-2 m)},\\
	& m_{33}= 0, \qquad m_{34}=\Delta _\theta( \pi/2 )=1,\\
	& m_{43}= \frac{ \Theta''(\pi/2 )}{2 \Delta_\theta (\pi/2)}= a^2 \left(E^2-m_0^2\right)-L^2.
\end{align}
Using the general formula \eqref{eq:LyaCO}, the Lyapunov exponents for a circular orbit in Kerr spacetime are
\begin{align}
	&\lambda_1=-\sqrt{\frac{\mathcal{R}''(r_c)}{2}}=-\sqrt{a^2 \left(E^2-m_0^2\right)+6 r_c \left(E^2 r_c+m m_0^2-m_0^2 r_c\right)-L^2},\\
	&\lambda_2=+\sqrt{\frac{\mathcal{R}''(r_c)}{2}}= \sqrt{a^2 \left(E^2-m_0^2\right)+6 r_c \left(E^2 r_c+m m_0^2-m_0^2 r_c\right)-L^2},\\
	&\lambda_3= - \sqrt{\frac{\Theta''(\pi/2)}{2}}= -\sqrt{a^2 \left(E^2-m_0^2\right)-L^2},\\
	&\lambda_4= + \sqrt{\frac{\Theta''(\pi/2)}{2}}=\sqrt{a^2 \left(E^2-m_0^2\right)-L^2}. 
\end{align}
In the limit $a\to 0$, the above Lyapunov exponents reduce to the corresponding results for the Schwarzschild black hole. Here, $E$ and $L$ denote the energy and angular momentum of the circular orbit, respectively, and are determined by the conditions $\mathcal{R}(r_c)=0$ and $\mathcal{R}'(r_c)=0$.

\subsubsection{Timelike case}

Let us consider a timelike circular orbit with $m_0=1$ in Kerr spacetime. The radial effective potential from \eqref{eq:kerrR} becomes
\begin{align}
	\mathcal{R}(r)&=\left(E \left(a^2+r^2\right)-a L\right)^2-\left(a^2-2 m r+r^2\right) \left(\Chat+m_0^2 r^2\right) \nonumber\\
	&=\left(E \left(a^2+r^2\right)-a L\right)^2-\left(a^2-2 m r+r^2\right) \left((L-a E)^2+r^2\right).
\end{align}
From $\mathcal{R}(r_c)=0$ and $\mathcal{R}'(r_c)=0$, we can obtain the energy and angular momentum at $r=r_c$
\begin{align}
	&E_1^2= \frac{\alpha+2 \sqrt{\beta}}{\eta}, \qquad L_1^2=\frac{\gamma+2 \left(a^2+3 r^2\right) \sqrt{\beta}}{\eta},\\
	&E_2^2= \frac{\alpha-2 \sqrt{\beta}}{\eta}, \qquad L_2^2=\frac{\gamma-2 \left(a^2+3 r^2\right) \sqrt{\beta}}{\eta},
\end{align}
with 
\begin{align}
	&\alpha=a^2 m  (5 m-3 r)r^2-12 m^3 r^3+16 m^2 r^4-7 m r^5+r^6,\\
	&\beta=m^3 r^3 \left(a^3+a r (r-2 m)\right)^2,\\
	&\gamma=m r^2 \left(a^4 (5 m+r)+2 a^2 r \left(-6 m^2+3 m r+r^2\right)+r^4 (r-3 m)\right),\\
	&\eta=r^3 \left(r (r-3 m)^2-4 a^2 m\right).
\end{align}
Note that $\mathcal{R}(r_c)=0$ and $\mathcal{R}'(r_c)=0$ yield two distinct sets of solutions for the energy and angular momentum, $(E_1,L_1)$ and $(E_2, L_2)$. These correspond to the prograde and retrograde circular orbits, respectively, reflecting the influence of the black hole spin.

Using the solution $(E_1,L_1)$, the Lyapunov exponents are
\begin{align}
	\lambda _1=- \lambda _2&= -\sqrt{\frac{a^2 (\alpha -\eta )-\gamma +6 r \left(\eta  m+r \left(\alpha +\sqrt{\beta }-\eta \right)\right)}{\eta }},\\
	\lambda _3=-\lambda_4&=-\sqrt{\frac{a^2 (\alpha -\eta )-\gamma -6 \sqrt{\beta } r^2}{\eta }}.
\end{align}
Using the other solution $(E_2,L_2)$, the corresponding Lyapunov exponents become
\begin{align}
	\lambda _1=-\lambda _2&=-\sqrt{\frac{a^2 (\alpha -\eta )-\gamma +6 r \left(\eta  m+r \left(\alpha -\sqrt{\beta }-\eta \right)\right)}{\eta }},\\
	\lambda _3=-\lambda _4&=-\sqrt{\frac{a^2 (\alpha -\eta )-\gamma +6 \sqrt{\beta } r^2}{\eta }}.
\end{align}
For timelike circular orbits in Kerr spacetime, the two branches yield distinct sets of Lyapunov exponents corresponding to the prograde and retrograde orbits, respectively. This differs from the Schwarzschild case.

\begin{figure}[htb]
	\centering
	\subfigure[]{\includegraphics[width=6cm]{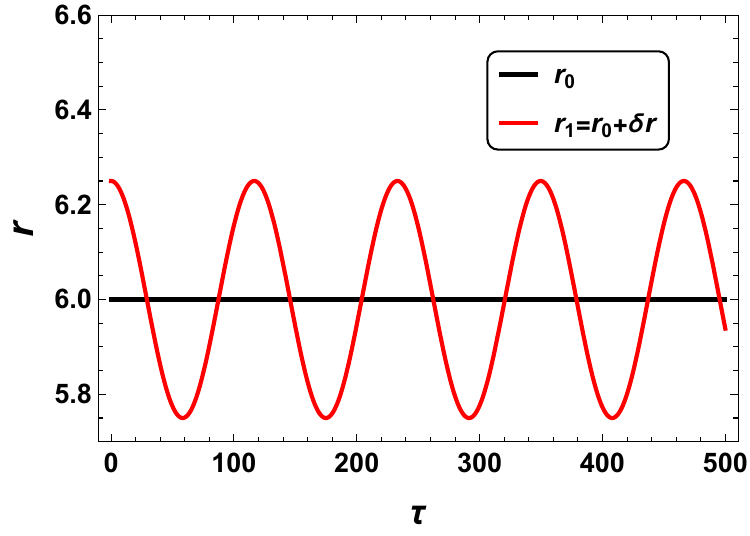}}\hspace{1cm}
	\subfigure[]{\includegraphics[width=6cm]{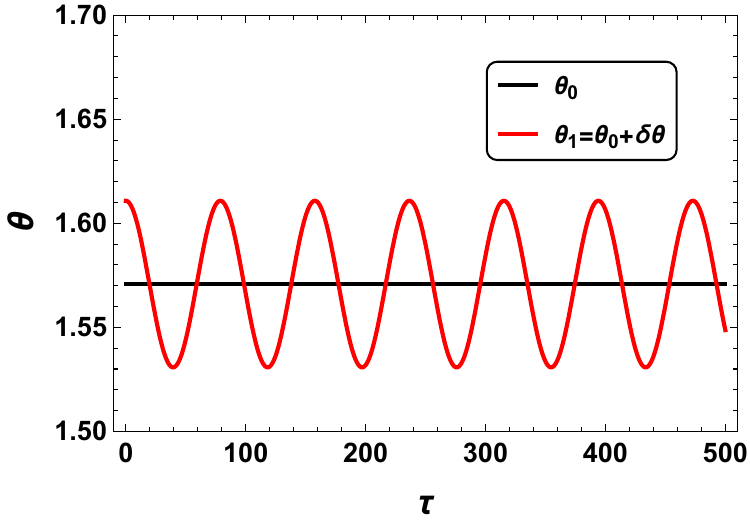}}
	\caption{Timelike circular orbit and its perturbed counterpart in Kerr spacetime. (a) Radial motion with $r_0=6$ and $r_1(\tau)=r_0+\delta r(\tau )$. (b) Polar motion with $\theta_0= \frac{\pi}{2}$ and $\theta_1(\tau )=\theta_0+\delta\theta(\tau )$. The initial perturbations are chosen as $\delta r(0)=0.2$ and $\delta\theta(0)=0.04$.}
	\label{fig:kerrTCO}
\end{figure}

\begin{figure}[htb]
	\centering
	\subfigure{\includegraphics[width=5cm]{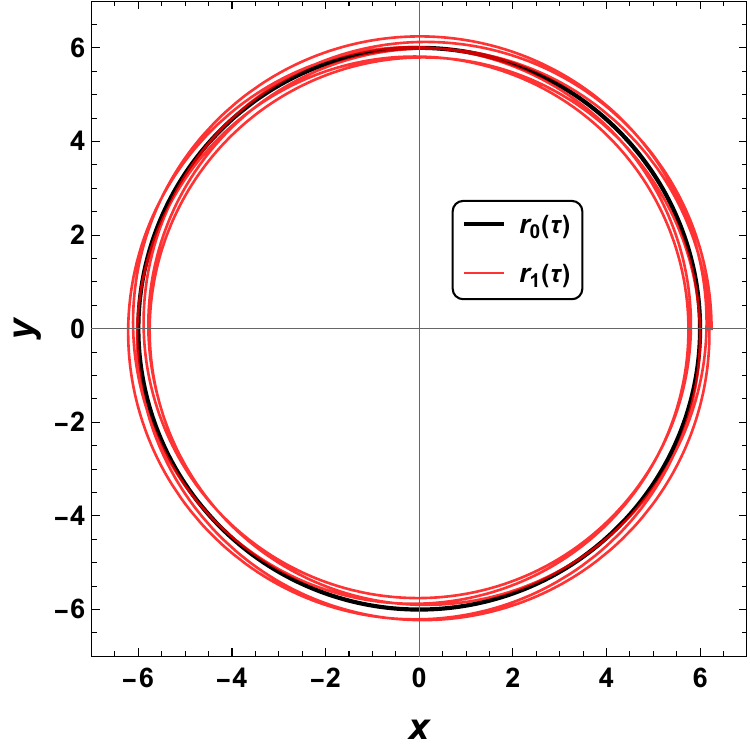}}\hspace{1cm}
	\subfigure{\includegraphics[width=5cm]{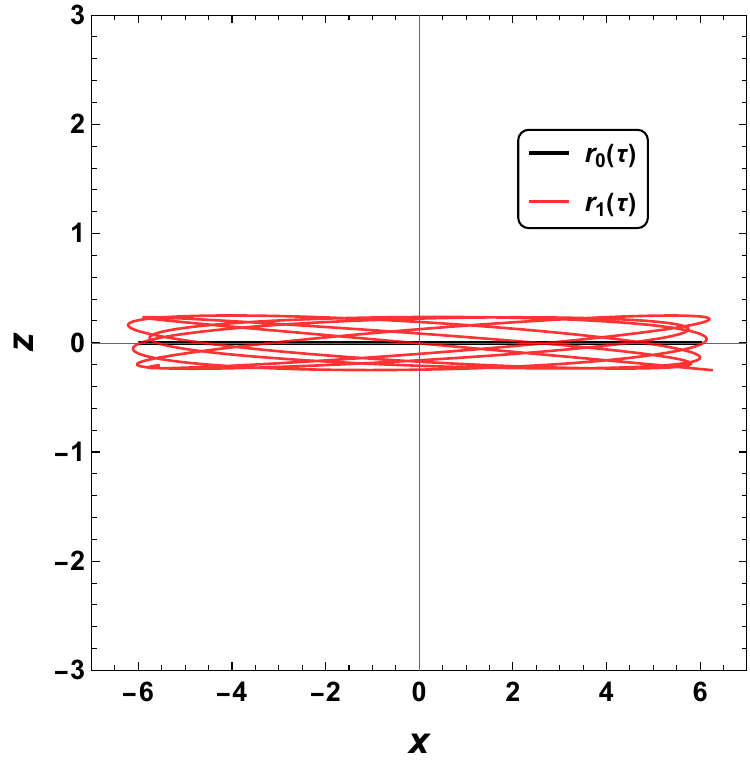}}
	\caption{Projections of a timelike circular orbit and its perturbed orbit onto the $(x,y)$ and $(x,z)$ planes in Kerr spacetime. Left: $(x,y)$ plane. Right: $(x,z)$ plane.} 
	\label{fig:kerrTCOxyz}
\end{figure}

To illustrate the dynamics, we consider the parameter choice $m=1$, $r_0=6$, and $\theta_0=\pi/2$. The trajectories $r_1(\tau)$ and $\theta_1(\tau)$ are obtained by numerically solving Hamilton's equations of motion and are shown in figure~\ref{fig:kerrTCO}. Furthermore, the projections of these trajectories onto the $(x,y)$ and $(x,z)$ planes are shown in figure~\ref{fig:kerrTCOxyz}. The trajectories in both radial and polar motions are stable.

\subsubsection{Null case}

For the null circular orbits with $m_0=0$ in Kerr spacetime, the radial effective potential from \eqref{eq:kerrR} is 
\begin{align}
	\mathcal{R}(r) =\left(E \left(a^2+r^2\right)-a L\right)^2-(L-a E)^2 \left(a^2-2 m r+r^2\right).
\end{align}
From $\mathcal{R}(r_c)=0$, one can obtain the relation between $E$ and $L$
\begin{align}
	&L_1= \frac{E \left(2 a m-r_c \sqrt{a^2+r_c (r_c-2 m)}\right)}{2 m-r_c},\\
	&L_2= \frac{E \left(2 a m+r_c \sqrt{a^2+r_c (r_c-2 m)}\right)}{2 m-r_c}.
\end{align}
Next, using $\mathcal{R}'(r_c)=0$, we obtain an equation for the radius $r_c$ of the null circular orbit
\begin{align}\label{eq:kerrNullcubic}
	r \left(4 a^2 E^2-4 a E L-2 (L-a E)^2\right)+2 m (L-a E)^2+4 E^2 r^3=0.
\end{align}
Since this is a cubic equation, its roots can be obtained analytically using the general solution. It should be noted that only the solutions satisfying $r>r_h=m+\sqrt{m^2-a^2}$ are physically relevant, as they lie outside the event horizon of the Kerr black hole.

Using the above angular momentum $L_1$, the Lyapunov exponents take the form
\begin{align}
	\lambda _1&=-\lambda _2 \nonumber \\
	& =-E \bigg[\frac{ 4 a m r_c \sqrt{a^2+r_c (r_c-2 m)}-4 a^2 m r_c+r_c^2 (12 m-5 r_c) (2 m-r_c)}{(r_c-2 m)^2}\bigg]^{1/2},\\[2mm]
	\lambda _3&=-\lambda_4=-E \bigg[a^2-\frac{(r_c \sqrt{a^2+r_c (r_c-2 m)}-2 a m)^2}{(r_c-2 m)^2}\bigg]^{1/2}.
\end{align}
Using the other solution $L_2$, the Lyapunov exponents are
\begin{align}
	\lambda _1&=-\lambda _2 \nonumber\\
	&=-E \bigg[\frac{ -4 a m r_c \sqrt{a^2+r_c (r_c-2 m)}-4 a^2 m r_c+r_c^2 (12 m-5 r_c) (2 m-r_c)}{(r_c-2 m)^2}\bigg]^{1/2},\\
	\lambda _3&=-\lambda_4=-E \bigg[a^2-\frac{(r_c \sqrt{a^2+r_c (r_c-2 m)}+2 a m)^2}{(r_c-2 m)^2}\bigg]^{1/2}.
\end{align}
Note that two distinct Lyapunov exponents are obtained, corresponding to the angular momenta $L_1$ and $L_2$, respectively. Furthermore, the null circular orbit radius $r_c$ appearing in these expressions must be determined by solving the cubic equation \eqref{eq:kerrNullcubic}.

\begin{figure}[htb]
	\centering
	\subfigure[]{\includegraphics[width=6cm]{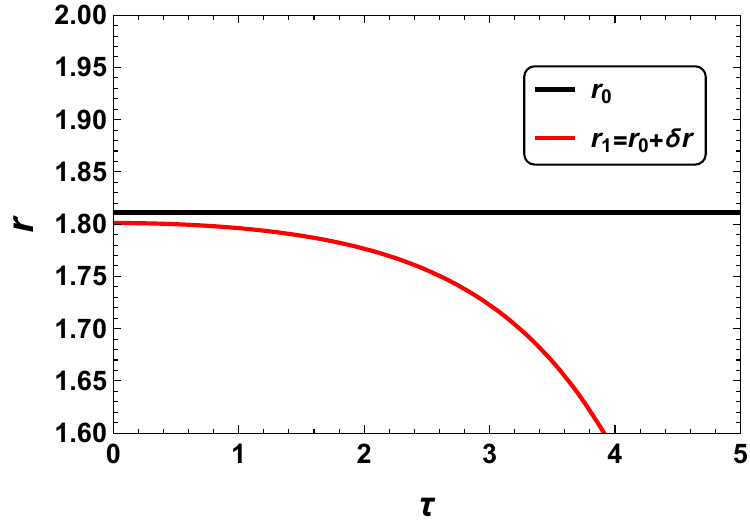}}\hspace{1cm}
	\subfigure[]{\includegraphics[width=6cm]{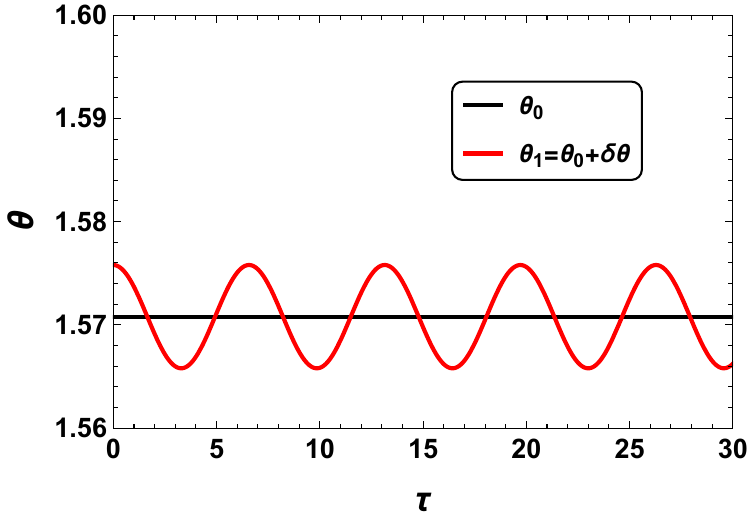}}
	\caption{Null circular orbit and its perturbed counterpart in Kerr spacetime. (a) Radial motion with $r_0=1.8110$ and $r_1(\tau)=r_0+\delta r(\tau )$. (b) Polar motion with $\theta_0= \frac{\pi}{2}$ and $\theta_1(\tau )=\theta_0+\delta\theta(\tau )$. The initial perturbations are chosen as $\delta r(0)=-0.01$ and $\delta\theta(0)=0.005$.}
	\label{fig:kerrNCO}
\end{figure}

\begin{figure}[htb]
	\centering
	\subfigure{\includegraphics[width=5cm]{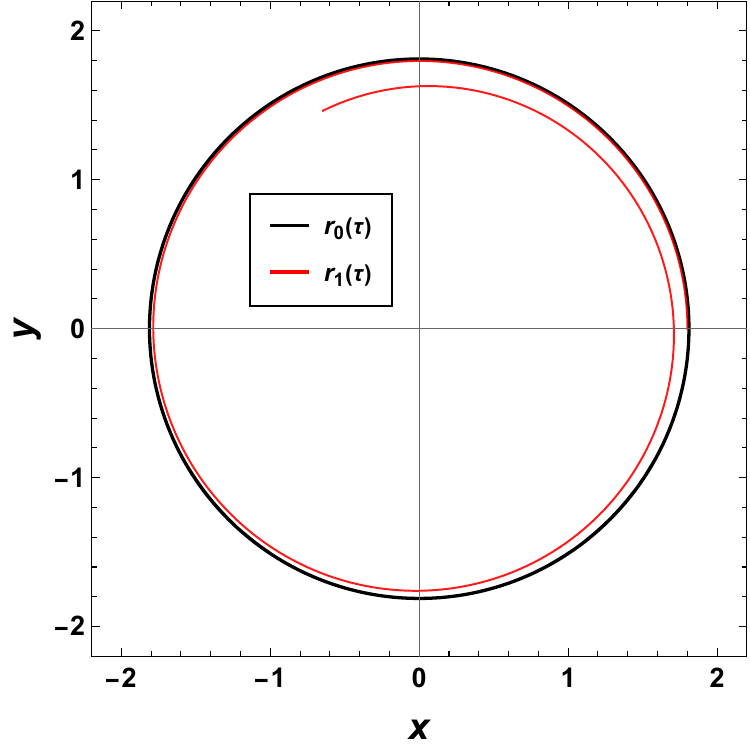}}\hspace{1cm}
	\subfigure{\includegraphics[width=5cm]{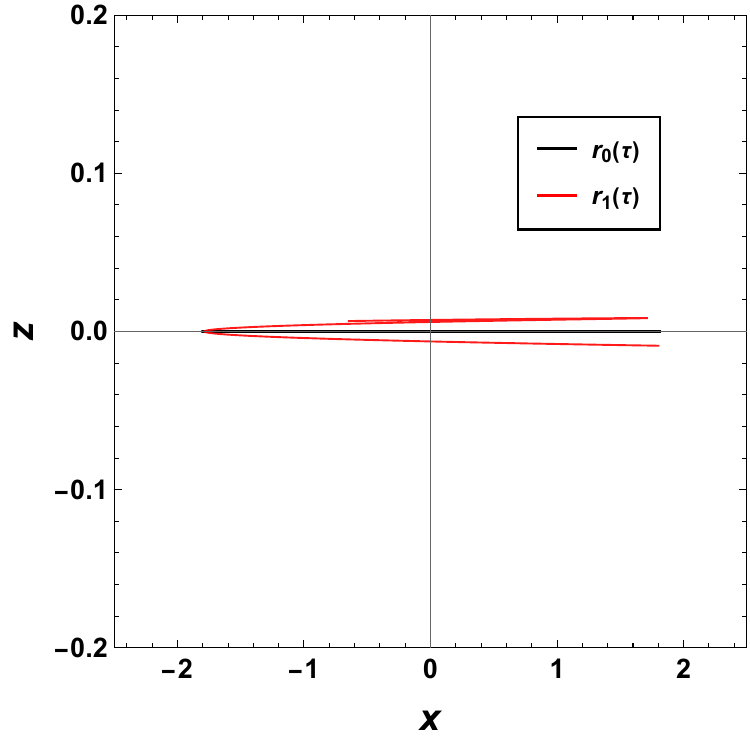}}
	\caption{Projections of a null circular orbit and its perturbed orbit onto the $(x,y)$ and $(x,z)$ planes in Kerr spacetime. Left: $(x,y)$ plane. Right: $(x,z)$ plane.} 
	\label{fig:kerrNCOxyz}
\end{figure}

As an illustrative example, we choose $m=1$, $a=0.8$, $r_0=1.8110$, and $\theta_0=\pi/2$. Hamilton's equations of motion are then solved numerically to obtain the trajectories $r_1(\tau)$ and $\theta_1(\tau)$, which are shown in figure~\ref{fig:kerrNCO}. These trajectories are projected onto the $(x,y)$ and $(x,z)$ planes, and the corresponding projections are shown in figure~\ref{fig:kerrNCOxyz}. We can find that the trajectory in radial direction is unstable, while in polar direction is stable.

\subsection{Lyapunov exponents for spherical orbits}

We now investigate spherical orbits in the Kerr spacetime. Such orbits are characterized by a constant radial coordinate $r=r_s$, whereas the polar angle $\theta$ varies within the range $\theta_{\min}\leq\theta\leq\theta_{\max}$. The conditions for the radial and polar motions can be expressed as
\begin{align}
	&\mathcal{R}(r_s)=0,  \quad \mathcal{R}'(r_s)=0,\\
	&\Theta(\theta_{\min})= \Theta(\theta_{\max})=0. 
\end{align}
From $\mathcal{R}(r_s)=0$ and $\Theta(\theta_{\min} )=0$, it follows that
\begin{align}
	&p_r= \sqrt{\mathcal{R}(r_s)}/\Delta_r(r_s)=0,\\
	&\Chat\; = a^2 m_0^2 \cos ^2(\theta_0)+(L \csc (\theta_0)-a E \sin (\theta_0))^2,
\end{align}
where we denote $\theta_{\min }=\theta_0$. Imposing the above conditions, the matrix elements of the linear stability matrix become
\begin{align}
	& m_{11}=0, \qquad m_{12}= \Delta _r(r_s)=r_s^2-2m r_s+a^2,\\[1mm]
	&m_{22}=\frac{\mathcal{R}''(r_s)}{2 \Delta_r(r_s)}=\frac{a^2 \left(2 E^2-m_0^2\right)-2 a E L-\Chat+6 r_s \left(E^2 r_s+m m_0^2-m_0^2 r_s\right)}{a^2+r_s (r_s-2 m)},\\[1mm]
	& m_{33}= 0 , \qquad m_{34}=1,\\[1mm]
	& m_{43}=a^2 \left(m_0^2-E^2\right) \cos (2 \theta )-L^2 (\cos (2 \theta )+2) \csc ^4(\theta ).
\end{align}
According to the general formula \eqref{eq:LyaSO}, the Lyapunov exponents for spherical orbits in Kerr spacetime are 
\begin{align}
	\lambda_1=&-\sqrt{\mathcal{R}''(r)/2} = -\big[a^2 \left(2 E^2-m_0^2\right) -a^2 \left(E^2 \sin ^2\theta_0+m_0^2 \cos ^2\theta_0\right) -L^2 \csc ^2\theta_0 \nonumber\\
	&\qquad +6 r_s \left(E^2 r_s+m_0^2 (m-r_s)\right)\big]^{1/2},\\[1mm]
	\lambda_2=&+\sqrt{\mathcal{R}''(r)/2} =+\big[a^2 \left(2 E^2-m_0^2\right) -a^2 \left(E^2 \sin ^2\theta_0+m_0^2 \cos ^2\theta_0\right) -L^2 \csc ^2\theta_0 \nonumber\\
	&\qquad +6 r_s \left(E^2 r_s+m_0^2 (m-r_s)\right)\big]^{1/2},\\[1mm]
	\lambda_3=&-\frac{1}{4} i \csc ^2(\theta ) \big[4 a^2 \left(m_0^2-E^2\right)+\cos (2 \theta ) \left(7 a^2 (E-m_0) (E+m_0)+16 L^2\right) \nonumber\\
	& \qquad +a^2 (E-m_0) (E+m_0) (\cos (6 \theta )-4 \cos (4 \theta ))+32 L^2\big]^{1/2},\\[1mm]
	\lambda_4=&+\frac{1}{4} i \csc ^2(\theta ) \big[4 a^2 \left(m_0^2-E^2\right)+\cos (2 \theta ) \left(7 a^2 (E-m_0) (E+m_0)+16 L^2\right) \nonumber\\
	& \qquad +a^2 (E-m_0) (E+m_0) (\cos (6 \theta )-4 \cos (4 \theta ))+32 L^2\big]^{1/2}.
\end{align}
In $\lambda_1$ and $\lambda_2$, we have used $\mathcal{C}=\hat{\mathcal{C}}$. The quantities $E$ and $L$ correspond to the conserved energy and angular momentum of the spherical orbit, respectively, and are obtained from the conditions $\mathcal{R}(r_s)=0$ and $\mathcal{R}'(r_s)=0$.

\subsubsection{Timelike case}

Let us consider timelike spherical orbits with $m_0=1$. The radial effective potential from \eqref{eq:kerrR} becomes
\begin{align}
	\mathcal{R}(r)&=\left(E \left(a^2+r^2\right)-a L\right)^2-\left(a^2-2 m r+r^2\right) \left(\Chat+m_0^2 r^2\right) \nonumber\\
	&=\left(E \left(a^2+r^2\right)-a L\right)^2-\left(a^2-2 m r+r^2\right) \big[r^2+a^2 m_0^2 \cos ^2\theta_0 \nonumber \\
	& \qquad +(L \csc \theta_0-a E \sin \theta_0)^2\big].
\end{align}
The conditions $\mathcal{R}(r_s)=0$ and $\mathcal{R}'(r_s)=0$ can be expressed as 
\begin{equation}\label{eq:EL2}
	\begin{aligned}
		&a_0 L^2+b_0 LE+c_0 E^2+d_0 =0,\\
		&a_1 L^2+b_1 LE+c_1 E^2+d_1 =0,
	\end{aligned}
\end{equation}
with 
\begin{align*}
	&a_0=a^2-\csc ^2(\theta_0) \left(a^2+r_s (r_s-2 m)\right),\\
	& b_0=-4 a m r_s,\\
	&c_0=\left(a^2+r_s^2\right)^2-a^2 \sin ^2(\theta_0) \left(a^2+r_s (r_s-2 m)\right),\\
	&d_0=-\left(\left(a^2+r_s (r_s-2 m)\right) \left(a^2 m_0^2 \cos ^2(\theta_0)+r_s^2\right)\right),\\
	&a_1=2 \csc ^2(\theta_0) (m-r_s),\\
	&b_1=-4 a m,\\
	&c_1=2 a^2 \sin ^2(\theta_0) (m-r_s)+4 r_s \left(a^2+r_s^2\right),\\
	&d_1=2 a^2 m_0^2 \cos ^2(\theta_0) (m-r_s)-2 r_s \left(a^2+r_s (2 r_s-3 m)\right).
\end{align*}
The energy and angular momentum can be obtained from \eqref{eq:EL2}
\begin{align}
	&E_1^2= \frac{\alpha+\sqrt{\beta}}{\eta}, \qquad L_1^2= \frac{\gamma+(b_1 c_0-b_0 c_1) \sqrt{\beta}}{(a_1 b_0-a_0 b_1) \eta },\\
	&E_2^2= \frac{\alpha-\sqrt{\beta}}{\eta}, \qquad L_2^2= \frac{\gamma-(b_1 c_0-b_0 c_1) \sqrt{\beta}}{(a_1 b_0-a_0 b_1) \eta },
\end{align}
with 
\begin{align*}
	\alpha=&-2 a_0^2 c_1 d_1+a_1 \left(2 a_0 (c_0 d_1+c_1 d_0) -b_0^2 d_1+b_0 b_1 d_0\right)+a_0 b_0 b_1 d_1 \\
	& -a_0 b_1^2 d_0-2 a_1^2 c_0 d_0, \\[1mm]
	\beta=&(a_1 b_0-a_0 b_1)^2 \big[2 d_0 d_1 (2 a_0 c_1+2 a_1 c_0-b_0 b_1)+d_1^2 \left(b_0^2-4 a_0 c_0\right) \\
	&+d_0^2 \left(b_1^2-4 a_1 c_1\right)\big], \\[1mm]
	\gamma = &2 a_0^2 b_1 c_1 (c_1 d_0-c_0 d_1)+a_1 \big[-b_0 \left(2 a_0 c_1 (c_1 d_0-c_0 d_1)+b_1^2 c_0 d_0\right) \\
	&+2 a_0 b_1 c_0 (c_0 d_1-c_1 d_0)+b_0^3 (-c_1) d_1+b_0^2 b_1 (c_0 d_1+c_1 d_0)\big] \\
	&+a_0 b_1 (b_1 c_0-b_0 c_1) (b_1 d_0-b_0 d_1)+2 a_1^2 b_0 c_0 (c_1 d_0-c_0 d_1),\\[1mm]
	\eta =&2 \left(a_1 \left(-2 a_0 c_0 c_1+b_0^2 c_1-b_0 b_1 c_0\right)+a_0 \left(a_0 c_1^2-b_0 b_1 c_1+b_1^2 c_0\right)+a_1^2 c_0^2\right).
\end{align*}
Equation~\eqref{eq:EL2} yields two branches of solutions, $(E_1,L_1)$ and $(E_2,L_2)$, for timelike spherical orbits. These branches are associated with prograde and retrograde motion, respectively, reflecting the influence of the black hole spin.

The Lyapunov exponents can then be written as
\begin{align}
	\lambda _1=-\lambda _2=&-\big[-a^2 \left(E^2 \sin ^2(\theta_0)+\cos ^2(\theta_0)\right)+a^2 \left(2 E^2-1\right) \nonumber\\
	&+6 r \left(\left(E^2-1\right) r+m\right)-L^2 \csc ^2(\theta_0)]^{1/2},\\[1mm]
	\lambda _3=-\lambda _4=&-\frac{1}{4} i \csc ^2(\theta ) \big[a^2 \left(E^2-1\right) (\cos (6 \theta )-4 \cos (4 \theta )) \nonumber\\
	&+\cos (2 \theta ) \left(7 a^2 \left(E^2-1\right)+16 L^2\right) -4 a^2 \left(E^2-1\right)+32 L^2\big]^{1/2}.
\end{align}
Since $E$ and $L$ admit two solution branches, $(E_1,L_1)$ and $(E_2,L_2)$, the above expressions give rise to two distinct sets of Lyapunov exponents. These are associated with the prograde and retrograde spherical orbits, respectively.

\begin{figure}[htb]
	\centering
	\subfigure[]{\includegraphics[width=6cm]{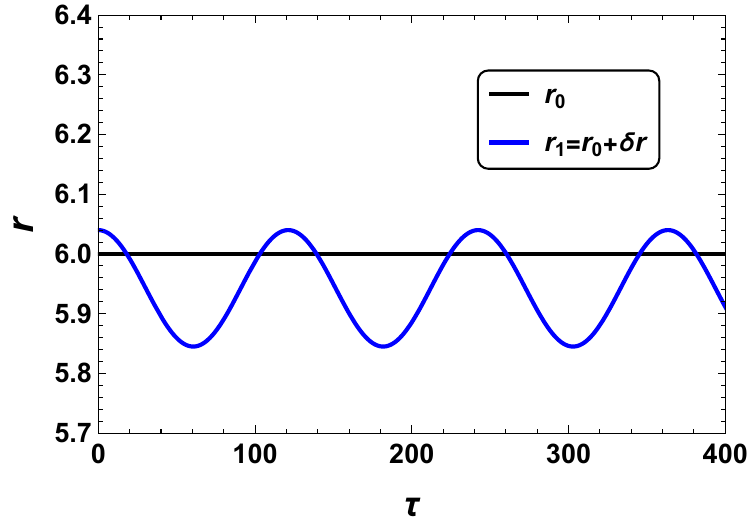}}\hspace{1cm}
	\subfigure[]{\includegraphics[width=6cm]{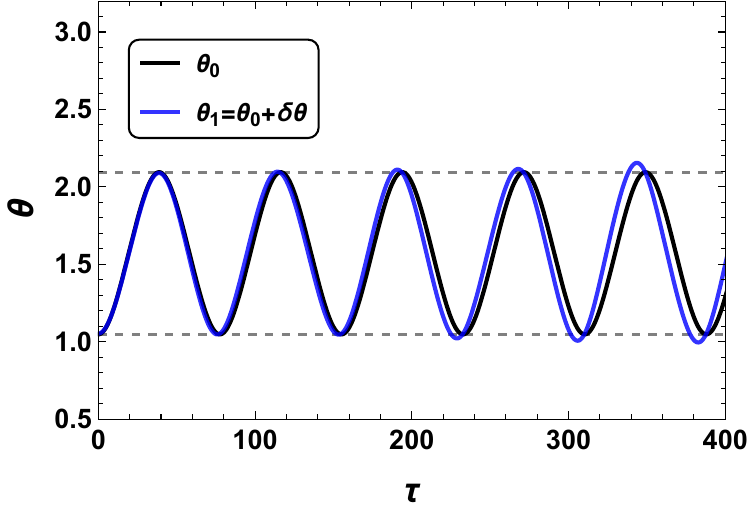}}
	\caption{Timelike spherical orbit and its perturbed counterpart in Kerr spacetime. (a) Radial motion with $r_0=6$ and $r_1(\tau )=r_0+\delta r(\tau )$. (b) Polar motion with $\theta(\tau ) \in [ \frac{\pi}{3},\frac{2 \pi}{3}]$ and $\theta_1(\tau )=\theta (\tau )+\delta\theta(\tau)$. The initial perturbations are chosen as $\delta r(0)=0.04$ and $\delta\theta(0)=0.005$.}
	\label{fig:kerrTSO}
\end{figure}

\begin{figure}[htb]
	\centering
	\subfigure{\includegraphics[width=5cm]{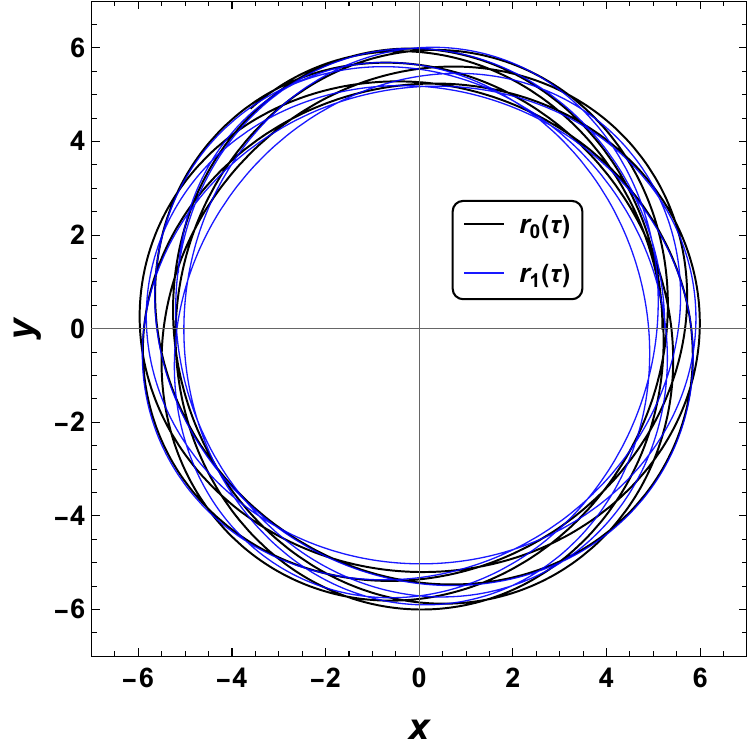}}\hspace{1cm}
	\subfigure{\includegraphics[width=5cm]{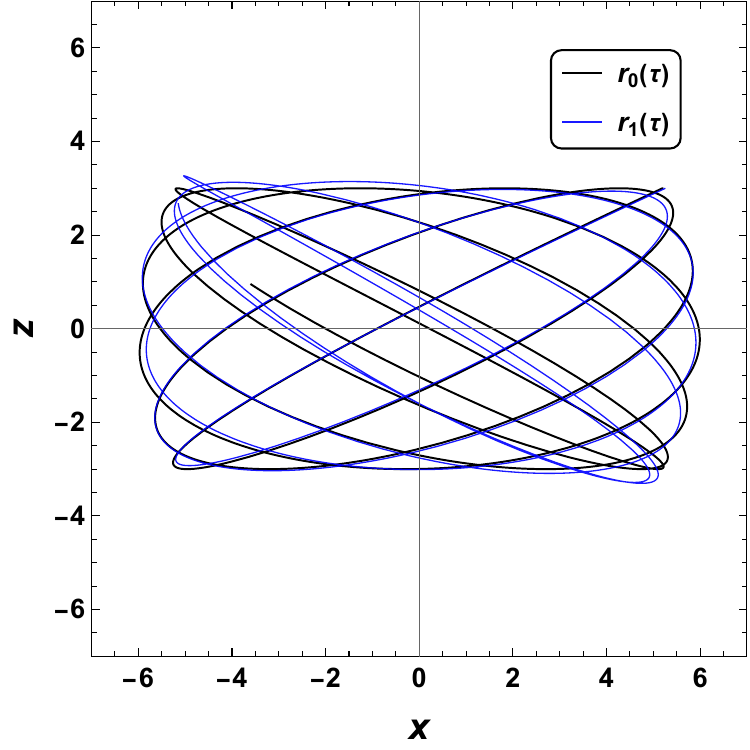}}
	\caption{Projections of a timelike spherical orbit and its perturbed orbit onto the $(x,y)$ and $(x,z)$ planes in Kerr spacetime. Left: $(x,y)$ plane. Right: $(x,z)$ plane.} 
	\label{fig:kerrTSOxyz}
\end{figure}

We now present an explicit example of a spherical orbit with $m=1$, $a=0.8$, $r_0=6$, and $\theta_0=\frac{\pi}{3}$. The corresponding polar motion oscillates within the range $\theta\in[\frac{\pi}{3},\frac{2\pi}{3}]$. Numerically solving Hamilton's equations of motion yields the trajectories $r_1(\tau)$ and $\theta_1(\tau)$ shown in figure~\ref{fig:kerrTSO}. For visualization purposes, we further project the orbit onto the $(x,y)$ and $(x,z)$ planes. The resulting projected trajectories are shown in figure~\ref{fig:kerrTSOxyz}. Obviously, the trajectory is stable against the linear perturbation, also indicated in the Lyapunov exponents.

\subsubsection{Null case}

Let us consider null spherical orbits in Kerr spacetime by setting $m_0=0$.
The radial effective potential from \eqref{eq:kerrR} becomes 
\begin{align}
	\mathcal{R}(r) =\left(E \left(a^2+r^2\right)-a L\right)^2-\left(a^2-2 m r+r^2\right) (L \csc (\theta_0)-a E \sin (\theta_0))^2.
\end{align}
From $\mathcal{R}(r_s)=0$, the relation between $L$ and $E$ can be obtained as
\begin{equation}\label{eq:L2}
	\begin{aligned}
		L_1&= \frac{4 a E m r_s-E \csc (\theta_0) \sqrt{a^2-2 m r_s+r_s^2} \left(a^2 \cos (2 \theta_0)+a^2+2 r_s^2\right)}{2 \left(a^2 \left(-\csc ^2(\theta_0)\right)+a^2+2 m r_s \csc ^2(\theta_0)-r_s^2 \csc ^2(\theta_0)\right)},\\[2mm]
		L_2&= \frac{E \csc (\theta_0) \sqrt{a^2-2 m r_s+r_s^2} \left(a^2 \cos (2 \theta_0)+a^2+2 r_s^2\right)+4 a E m r_s}{2 \left(a^2 \left(-\csc ^2(\theta_0)\right)+a^2+2 m r_s \csc ^2(\theta_0)-r_s^2 \csc ^2(\theta_0)\right)}.
	\end{aligned}
\end{equation}
In addition, the condition $\mathcal{R}'(r_s)=0$ yields a cubic equation for null spherical orbits
\begin{align}\label{eq:rs3}
	&4 E^2 r_s^3+r_s \left(4 a^2 E^2-2 (L \csc (\theta_0)-a E \sin (\theta_0))^2-4 a E L\right) \nonumber\\
	& \qquad +2 m (L \csc (\theta_0)-a E \sin (\theta_0))^2=0.
\end{align}
The cubic equation can be solved analytically in closed form. However, only the roots satisfying $r>r_h=m+\sqrt{m^2-a^2}$ correspond to spherical orbits located outside the Kerr event horizon and are therefore physically admissible.

The Lyapunov exponents for null spherical orbits take the form
\begin{align}
	\lambda _1=-\lambda _2&=-\sqrt{E^2 \left(a^2 \left(-\sin ^2(\theta_0)\right)+2 a^2+6 r_s^2\right)-L^2 \csc ^2(\theta_0)},\label{LyakerrNSOr} \\[2mm]
	\lambda _3=-\lambda _4&=-\frac{1}{4} i \csc ^2(\theta ) \big[a^2 E^2 (\cos (6 \theta )-4 \cos (4 \theta ))+\cos (2 \theta ) \left(7 a^2 E^2+16 L^2\right) \nonumber\\
	& \quad -4 a^2 E^2+32 L^2\big]^{1/2}. \label{LyakerrNSOth}
\end{align}
Since the angular momentum from \eqref{eq:L2} admits two branches, $L_1$ and $L_2$, and the radius of the null spherical orbit is determined by the cubic equation \eqref{eq:rs3}, the above expressions give rise to two distinct sets of Lyapunov exponents. These are associated with the prograde and retrograde spherical orbit branches, respectively.

\begin{figure}[htb]
	\centering
	\subfigure{\includegraphics[width=5cm]{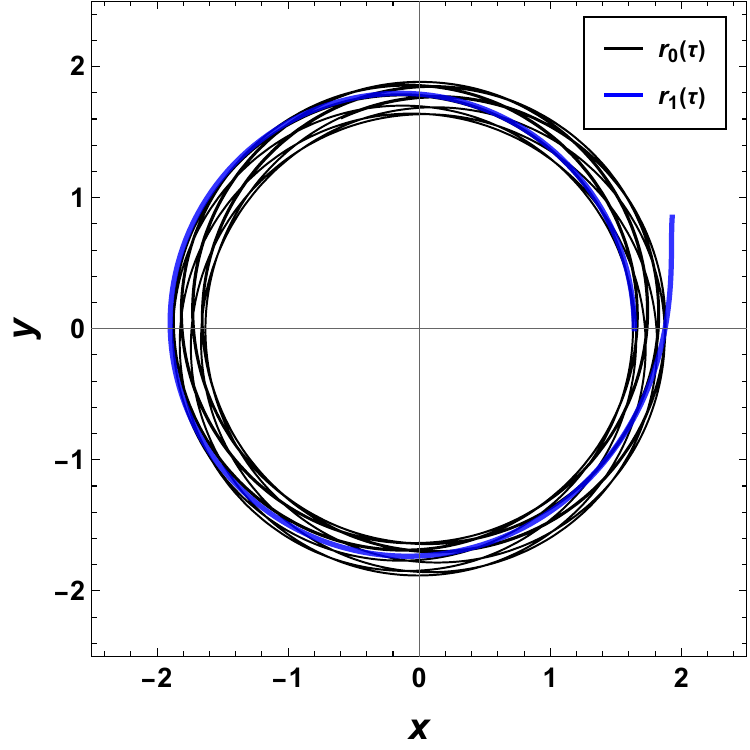}}\hspace{1cm}
	\subfigure{\includegraphics[width=5cm]{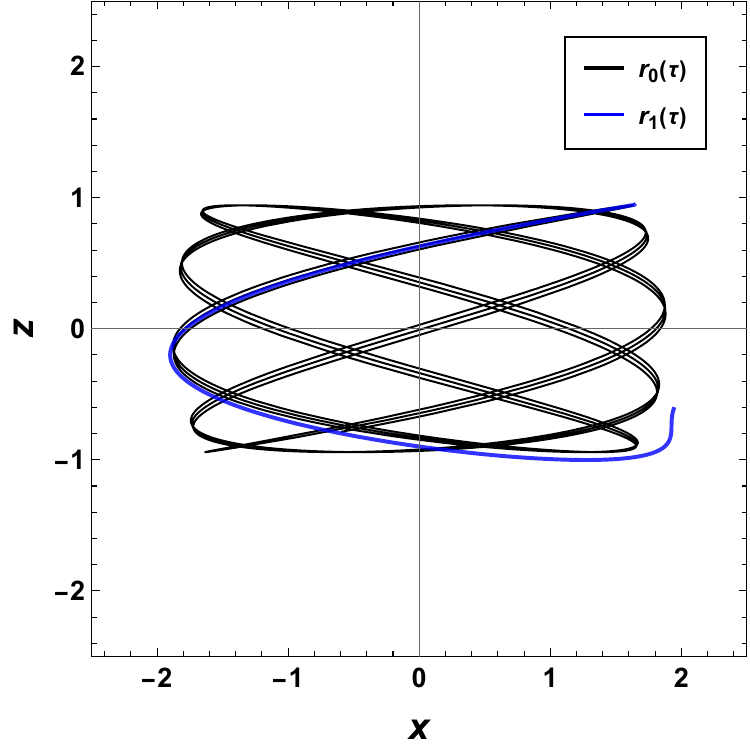}}
	\caption{Projections of a null spherical orbit and its perturbed orbit onto the $(x,y)$ and $(x,z)$ planes in Kerr spacetime. Left: $(x,y)$ plane. Right: $(x,z)$ plane. Here, the null spherical orbit is described by $r_0=1.8832$ and $\theta(\tau)\in[\frac{\pi}{3},\frac{2\pi}{3}]$. The perturbed orbit is described by $r_1(\tau)=r_0+\delta r(\tau)$ and $\theta_1(\tau)=\theta(\tau)+\delta\theta(\tau)$.}
	\label{fig:kerrNSOxyz}
\end{figure}

We consider a representative null spherical orbit with $m=1$, $a=0.8$, $r_0=1.8832$, and $\theta_0=\frac{\pi}{3}$, for which the polar angle oscillates within the range $\theta\in[\frac{\pi}{3},\frac{2\pi}{3}]$. The initial perturbations are chosen as $\delta r(0)=0.01$ and $\delta\theta(0)=0.001$. The resulting trajectories $r_1(\tau)$ and $\theta_1(\tau)$ are similar to those in the Schwarzschild case shown in figure~\ref{fig:schNSO}. Consequently, we present only the projections of the orbit onto the $(x,y)$ and $(x,z)$ planes, as shown in figure~\ref{fig:kerrNSOxyz}.

As time evolves, the perturbed radial trajectory $r_1(\tau)$ gradually deviates from the background orbit $r_0(\tau)$, indicating the instability of the radial perturbation, in agreement with the Lyapunov exponent given in \eqref{LyakerrNSOr}. In contrast, the polar trajectory $\theta_1(\tau)$ remains close to the reference trajectory $\theta_0(\tau)$ at early times, indicating the stability of the polar perturbation, consistent with the Lyapunov exponents given in \eqref{LyakerrNSOth}.

\section{Conclusion}\label{sec:conclusion}

In the present work, we studied the linear stability matrix and Lyapunov exponents in general stationary spacetimes. Our main results are summarized as follows.

First, for a general stationary spacetime with metric $g_{\mu\nu}=g_{\mu\nu}(r,\theta)$, we showed that the rank of the linear stability matrix satisfies $\mathrm{rank}(M_{LP})=4$. By analyzing Hamilton's equations of motion for a test particle, we obtained six nontrivial equations of motion in addition to $\dot{p}_t=0$ and $\dot{p}_\phi=0$. These equations determine the time evolution of the particle trajectory. The corresponding six-dimensional phase space is spanned by the basis $(t,r,\theta,\phi,p_r,p_\theta)$, and the linear stability matrix is given by the Jacobian matrix with respect to these variables, as shown in Eq.~\eqref{eq:LSmat}. However, the fact that its rank is four rather than six implies that only four independent linear perturbation equations exist. Consequently, we chose $(r,\theta,p_r,p_\theta)$ as the phase-space basis for the analysis of Lyapunov exponents. The coordinates $t$ and $\phi$ were excluded because they are cyclic coordinates and therefore do not contribute to the dynamical perturbations of the system.

Second, we showed that the $4\times4$ linear stability matrix in the basis $(r,\theta,p_r,p_\theta)$ can be further reduced to a block-diagonal form when the stationary spacetime admits an additional rank-2 Killing tensor. The existence of such a Killing tensor renders the Hamilton--Jacobi equation separable and leads to the decoupling of the radial and polar equations of motion. In this case, the linear stability matrix takes the form
\begin{equation*}
	M_{LP}=
	\renewcommand{\arraystretch}{1.}
	\left[
	\begin{array}{ >{\centering\arraybackslash}m{0.75 cm} >{\centering\arraybackslash}m{0.75cm}: >{\centering\arraybackslash}m{0.75 cm} >{\centering\arraybackslash}m{0.75cm} }
		$m_{11}$ & $m_{12}$ & 0 & 0 \\
		$m_{21}$ & 0 & 0 & 0 \\
		\noalign{\vskip 1mm}
		\hdashline[4pt/3pt]
		0 & 0 & $m_{33}$ & $m_{34}$ \\
		0 & 0 & $m_{43}$ & 0
	\end{array}
	\right],
\end{equation*}
with
\begin{align*}
	\lambda_{1,2} &=\frac{1}{2}\left( m_{11}\mp\sqrt{m_{11}^2+4m_{12}m_{21}} \right), \\ \lambda_{3,4}&=\frac{1}{2}\left( m_{33}\mp\sqrt{m_{33}^2+4m_{34}m_{43}} \right).
\end{align*}
The explicit expressions for the matrix elements are given in
Eq.~\eqref{eq:LPele}.

Third, for circular and spherical orbits in stationary spacetimes, the corresponding Lyapunov exponents can be obtained directly from the general formalism by imposing the appropriate constraint conditions. The physical interpretation of the Lyapunov exponents is straightforward. If $\lambda_{1,2}$ are purely imaginary, i.e., $\lambda_1=-\lambda_2=i\lambda$, the linear perturbation takes the form $\delta r\propto c_1 e^{i\lambda t}+c_2 e^{-i\lambda t} =a\cos(\lambda t)+b\sin(\lambda t)$, where $a=c_1+c_2$ and $b=i(c_2-c_1)$. In this case, the perturbed orbit $r(t)=r_0+\delta r(t)$ undergoes periodic oscillations around $r_0$, indicating that the background orbit is linearly stable. In contrast, when $\lambda_{1,2}$ are real and satisfy $\lambda_1=-\lambda_2=\lambda$, the perturbation $\delta r\propto c_1 e^{\lambda t}+c_2 e^{-\lambda t}$ grows or decays exponentially with time. Consequently, the background orbit is linearly unstable. An analogous interpretation applies to the Lyapunov exponents $\lambda_{3,4}$, which characterize perturbations in the $\theta$ direction.

Furthermore, we demonstrated the applicability of the formalism through several representative examples. For classical systems, we examined the one-dimensional (inverted) harmonic oscillator and the two-dimensional Kepler problem. For relativistic systems, we investigated particle motion in Schwarzschild and Kerr spacetimes. These examples consistently support the validity and effectiveness of the proposed formalism.

\acknowledgments

We thank Chao-Hui Wang and Shan-Ping Wu for helpful and stimulating discussions. This work was supported by the National Natural Science Foundation of China (Grants No. 12475055 and No. 12247101).

\appendix

\section{Coefficients in the linear perturbation equations} \label{appA}

\begin{align*}
	&\frac{\partial f_1}{\partial \theta} =L \frac{\partial g^{t\phi}}{\partial \theta}-E \frac{\partial g^{tt}}{\partial \theta}, \qquad \frac{\partial f_1}{\partial r}= L \frac{\partial g^{t \phi}}{\partial r} -E \frac{\partial g^{tt}}{\partial r}, \\
	&\frac{\partial f_2}{\partial \theta} = p_r  \frac{\partial g^{rr}}{\partial \theta}, \qquad \frac{\partial f_2}{\partial r} = p_r \frac{\partial g^{rr}}{\partial r},  \qquad \frac{\partial f_2}{\partial p_r } =g^{rr}, \\
	&\frac{\partial f_3}{\partial \theta} = p_\theta \frac{\partial g^{{\theta \theta}}}{\partial \theta}, \qquad \frac{\partial f_3}{\partial r} = p_\theta \frac{\partial g^{\theta \theta}}{\partial r}, \qquad \frac{\partial f_3}{\partial p_\theta}= g^{\theta\theta}, \\
	&\frac{\partial f_4}{\partial \theta} = L \frac{\partial g^{\phi \phi}}{\partial \theta} -E \frac{\partial g^{t \phi}}{\partial \theta},  \qquad \frac{\partial f_4}{\partial r}= L \frac{\partial g^{\phi \phi}}{\partial r} -E \frac{g^{t \phi}}{\partial r}.
\end{align*}
\begin{align*}
	&\frac{\partial f_5}{\partial r} = \frac{1}{2}  \left(-E^2  \frac{\partial^2 g^{tt}}{\partial r^2}+2 E L \frac{\partial^2 g^{t \phi}}{\partial^2 r} -p_\theta^2 \frac{\partial^2 g^{\theta \theta}}{\partial \theta^2} - p_r^2 \frac{\partial^2 g^{rr}}{\partial r^2} -L^2 \frac{\partial^2 g^{\phi \phi}}{\partial r^2} \right), \\
	&\frac{\partial f_5}{\partial \theta} =    \frac{1}{2}   \left(-E^2 \frac{\partial^2 g^{tt}}{\partial r \partial \theta} +2 E L \frac{\partial^2 g^{t \phi}}{\partial r \partial \theta} -p_\theta^2 \frac{\partial^2 g^{\theta \theta}}{\partial r \partial \theta} -p_r^2 \frac{\partial^2 g^{rr}}{\partial r \partial \theta} -L^2 \frac{\partial^2 g^{ \phi \phi}}{\partial r \partial \theta} \right), \\ 
	&\frac{\partial f_5}{\partial p_r} =- p_r \frac{\partial g^{rr}}{\partial r}, \qquad \frac{\partial f_5}{\partial p_\theta} = - p_\theta \frac{\partial g^{\theta \theta}}{ \partial \theta}.
\end{align*}
\begin{align*}
	&\frac{\partial f_6}{\partial_r}=   \frac{1}{2}  \left(-E^2 \frac{\partial^2 g^{tt}}{\partial r \partial \theta} +2 E L \frac{\partial^2 g^{t \phi}}{\partial r \partial \theta} -p_\theta^2 \frac{\partial^2 g^{\theta \theta}}{\partial r \partial \theta} -p_r^2 \frac{\partial g^{rr}}{\partial r \partial \theta} -L^2 \frac{\partial^2 g^{\phi \phi}}{\partial r \partial \theta} \right), \\ 
	&\frac{\partial f_6}{\partial_\theta}= \frac{1}{2}   \left(-E^2 \frac{\partial^2 g^{tt}}{\partial \theta^2} +2 E L \frac{\partial^2 g^{t \phi}}{\partial \theta^2} -p_\theta^2 \frac{\partial^2 g^{\theta \theta}}{\partial^2 \theta} -p_r^2 \frac{\partial^2 g^{rr}}{\partial \theta^2}-L^2 \frac{\partial^2 g^{ \phi \phi}}{\partial \theta^2} \right),\\
	&\frac{\partial f_6}{\partial p_r}=-p_r \frac{\partial g^{rr}}{\partial \theta}, \qquad \frac{\partial f_6}{\partial p_\theta}= - p_\theta \frac{\partial g^{\theta \theta}}{\partial \theta}.
\end{align*}

\end{document}